\documentclass[pdflatex,sn-nature]{sn-jnl}%

\usepackage{graphicx}%
\usepackage{placeins}%
\usepackage{multirow}%
\usepackage{amsmath,amssymb,amsfonts}%
\usepackage{amsthm}%
\usepackage{mathrsfs}%
\usepackage[title]{appendix}%
\usepackage{xcolor}%
\usepackage{textcomp}%
\usepackage{manyfoot}%
\usepackage{booktabs}%
\usepackage{algorithm}%
\usepackage{algorithmicx}%
\usepackage{algpseudocode}%
\usepackage{listings}%
\let\orcidlogo\undefined
\usepackage{orcidlink}%

\theoremstyle{thmstyleone}%

\theoremstyle{thmstyletwo}%

\theoremstyle{thmstylethree}%

\begin{document}

\title[Differentiable DFT-Based Inverse Materials Design]{A Differentiable DFT-Based Framework for Inverse Materials Design}

\author*[1]{\fnm{Kohei} \sur{Ishii}\,\orcidlink{0000-0002-6343-1647}}\email{kishii@issp.u-tokyo.ac.jp}

\author[1,2]{\fnm{Hisazumi} \sur{Akai}\,\orcidlink{0000-0002-3551-7443}}\email{akai@issp.u-tokyo.ac.jp}

\author[3]{\fnm{Tetsuya} \sur{Fukushima}\,\orcidlink{0000-0001-6301-6104}}\email{t.fukushima@aist.go.jp}

\author[3]{\fnm{Hikari} \sur{Shinya}\,\orcidlink{0000-0002-6257-2114}}\email{hikari.shinya@aist.go.jp}

\author[4]{\fnm{Hyuga} \sur{Hosoi}\,\orcidlink{0009-0004-9474-9414}}\email{hyuga\_hosoi@mail.toyota.co.jp}

\author[4]{\fnm{Noritsugu} \sur{Sakuma}\,\orcidlink{0000-0002-6073-3957}}\email{noritsugu\_sakuma@mail.toyota.co.jp}

\author*[1]{\fnm{Koji} \sur{Inui}\,\orcidlink{0000-0002-1749-5439}}\email{koji-inui@issp.u-tokyo.ac.jp}

\affil*[1]{\orgdiv{Institute for Solid State Physics}, \orgname{University of Tokyo}, \orgaddress{\city{Kashiwa}, \country{Japan}}}

\affil[2]{\orgdiv{Department of Precision Engineering, Graduate School of Engineering}, \orgname{The University of Osaka}, \orgaddress{\city{Suita}, \country{Japan}}}

\affil[3]{\orgdiv{Materials DX Research Center}, \orgname{National Institute of Advanced Industrial Science and Technology}, \orgaddress{\city{Tsukuba}, \country{Japan}}}

\affil[4]{\orgdiv{Advanced Materials Engineering Division}, \orgname{Toyota Motor Corporation}, \orgaddress{\city{Susono}, \country{Japan}}}

\abstract{Discovering solid-state materials with target properties remains a central challenge in computational materials science.
Existing data-driven approaches rely on training data and may generalize poorly outside the training distribution.
Here we develop a first-principles inverse-design framework that requires no training data.
It integrates reverse-mode automatic differentiation (AD) into the Korringa--Kohn--Rostoker method combined with the coherent potential approximation (KKR-CPA), in which elemental concentrations serve as continuous optimization variables.
Reverse-mode AD computes all concentration gradients in one backward pass, enabling optimization over dozens of candidate elements.
Any differentiable quantity computable within KKR-CPA can serve as an optimization objective.
We demonstrate this generality in permanent-magnet and half-metal design, identifying candidates such as (Nd$_{0.983}$Pr$_{0.017}$)(Co$_{0.72}$Fe$_{0.28}$)$_2$Fe$_3$ and FeMn(As$_{0.83}$P$_{0.17}$), respectively.
We also synthesized materials with five SmCo$_5$-based compositions selected during optimization for higher magnetization.
Their measured magnetization agreed closely with calculated values.
Our framework offers a first-principles route to designing materials with target properties.}

\maketitle

Discovering solid-state materials with desired functional properties is a longstanding goal of materials science.
First-principles calculations predict the properties of a given material, whereas inverse design seeks to identify materials with specified properties.
Even within a given crystal structure type, the choice of constituent elements and their proportions creates an enormous compositional search space.
Evaluating every candidate through first-principles calculations is computationally prohibitive.
This motivates automated inverse-design methods that efficiently search for compositions with the desired properties.

Several approaches seek to make this search more efficient.
High-throughput screening automates first-principles evaluations, but its cost grows with the number of candidates~\cite{Curtarolo2013}.
Data-driven approaches, including Bayesian optimization~\cite{Lookman2019} and generative models~\cite{SanchezLengeling2018}, use learned models to guide the search, but can struggle to generalize beyond the training distribution~\cite{Schmidt2019,Himanen2019}.
A complementary strategy is to update compositions using gradients obtained directly from first-principles calculations, without training a surrogate model~\cite{Zunger2018,vonLilienfeld2009,vonRudorff2020,Tahmassebpur2025}.
Making this strategy practical for multicomponent materials requires efficient evaluation of the target property's gradients with respect to many composition variables.

Automatic differentiation (AD)~\cite{Baydin2018,Griewank2008} computes gradients of differentiable objectives by applying the chain rule throughout a numerical calculation.
In reverse mode, it obtains the gradient of a scalar objective with respect to all input variables in a single backward pass, typically at a computational cost only a few times that of evaluating the objective.
Unlike finite differences, it requires neither a separate objective evaluation for each variable nor a perturbation step size.
These features make reverse-mode AD particularly well suited to inverse design problems with many design variables.

Applying AD to first-principles methods could therefore enable the inverse design of materials.
So far, however, gradient-based inverse design has been demonstrated mostly at the model level---for model Hamiltonians and quantum transport~\cite{Inui2023,Inui2024,Hirasaki2024}, for porous materials~\cite{Liu2023}, and for machine-learned interatomic potentials~\cite{Nam2025}---while differentiable first-principles methods, developed for quantum chemistry~\cite{TamayoMendoza2018,Kasim2021,Kasim2022,Zhang2022} and plane-wave density functional theory (DFT)~\cite{Schmitz2026}, target geometry or parameter optimization rather than composition.
Inverse design of solid-state composition from first principles has thus rarely been attempted.
The obstacle is composition itself: in plane-wave DFT and other supercell-based methods, the element on each lattice site is a discrete choice, so composition is not a differentiable parameter.
Supercell-based approaches, such as special quasi-random structures~\cite{Zunger1990}, allow only discrete compositions for a given cell size.
The virtual crystal approximation~\cite{Nordheim1931,Bellaiche2000} restores continuity by replacing the constituent atoms on a site with a single virtual atom. 
However, this averaging disregards the individuality of the constituent atoms, including their distinct electronic characteristics, and consequently neglects electron scattering from chemical disorder. 
This can lead to an inaccurate description of the alloy's electronic structure.

A first-principles framework in which compositions are intrinsically continuous is the Korringa--Kohn--Rostoker (KKR) method~\cite{Korringa1947,Kohn1954} combined with the coherent potential approximation (CPA)~\cite{Soven1967,Velicky1968,Gyorffy1972}, hereafter KKR-CPA~\cite{Akai1989,Ebert2011KKR,Faulkner2018}.
KKR is a Green's-function formulation of the Kohn--Sham equations, and CPA performs the configurational average over substitutional disorder analytically: each crystallographic site is occupied by multiple elements with continuous concentrations, while the distinct scattering of each element is retained.
KKR-CPA requires no supercell, and its cost is nearly independent of the number of elements mixed on each site.
Historically, KKR-CPA has been the standard first-principles method for impurities, dilute alloys, and substitutionally disordered systems---regimes in which composition is genuinely continuous in physical reality.
These continuous-composition predictions are quantitatively reliable: across disordered alloys, KKR-CPA reproduces measured electronic structure~\cite{Stocks1978}, composition-dependent magnetic moments and Curie temperatures~\cite{Takahashi2007}, and transport properties such as the anomalous Hall effect~\cite{Lowitzer2010}, establishing it as a practical, experimentally validated method.
KKR-CPA thus offers a mature first-principles framework in which compositions are intrinsically continuous and differentiable variables.

\begin{figure}[!htbp]
\centering
\includegraphics[height=4.8cm]{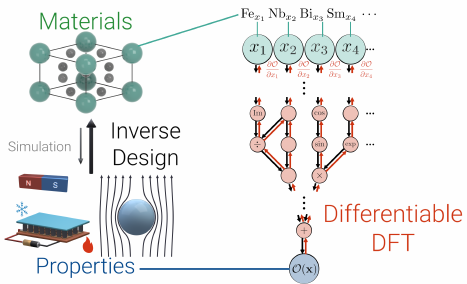}\hfill
\includegraphics[height=4.8cm]{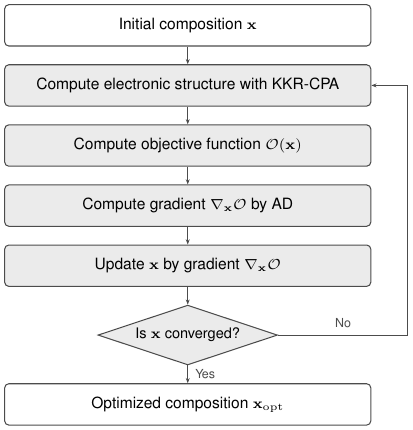}
\caption{\textbf{Gradient-based inverse materials design.}
\textbf{a}, Conceptual overview.
Left, the forward problem predicts material properties from a given composition, whereas inverse design seeks a composition that realizes a target property. Right, the Korringa--Kohn--Rostoker method combined with the coherent potential approximation (KKR-CPA) is represented as a computational graph connecting composition variables $\mathbf{x}$ to an objective $\mathcal{O}(\mathbf{x})$. Reverse-mode automatic differentiation (AD) propagates derivatives backward through this graph to obtain the composition gradient $\nabla_{\mathbf{x}}\mathcal{O}$.
\textbf{b}, Optimization workflow. Starting from an initial composition, the KKR-CPA solver (AkaiKKR) computes the self-consistent electronic structure. A differentiable implementation then evaluates the objective and obtains its composition gradient using reverse-mode AD. The optimizer uses this gradient to update the composition, and the cycle repeats until convergence.}
\label{fig:overview}
\end{figure}

Here we integrate reverse-mode AD into KKR-CPA, yielding gradients of an objective with respect to composition directly from the first-principles calculation and enabling gradient-based optimization (Fig.~\ref{fig:overview}).
The search is carried out within a prescribed crystal structure type, with the elements and their proportions on each crystallographic site as the design variables.
Fig.~\ref{fig:overview}a contrasts the conventional forward problem---predicting properties $\mathcal{O}(\mathbf{x})$ from a known composition---with the inverse problem solved here: finding compositions that yield target properties.
The right panel shows the conceptual KKR-CPA computational graph through which reverse-mode AD propagates gradients $\nabla_{\mathbf{x}}\mathcal{O}$ from output to input at a cost independent of the number of parameters.
The workflow (Fig.~\ref{fig:overview}b) implements this approach as an iterative loop: KKR-CPA forward evaluation produces $\mathcal{O}(\mathbf{x})$, reverse-mode AD provides $\nabla_{\mathbf{x}}\mathcal{O}$, and the composition is updated.
Iteration continues until convergence.

In this framework, dozens of elements can be considered and optimized all at once.
Every composition visited during the optimization is evaluated directly by a first-principles calculation, providing usable property predictions even at intermediate stages.
Because no machine-learned surrogate is involved, no training dataset is required, and the issue of extrapolation does not arise.
As the optimization proceeds, it drives the proportions of unneeded elements toward zero, thereby selecting the constituent elements.
Any differentiable objective computable within KKR-CPA can serve as the target, so the same method applies to a wide range of design goals.
By obtaining gradients for all composition variables in a single backward pass, the approach enables efficient inverse design in large compositional spaces.

To demonstrate the framework's generality, we consider two contrasting targets: magnetization and the shape of the density of states (DOS).
We first optimize the magnetization, $M$, motivated by permanent-magnet alloys, whose properties are sensitive to elemental substitution~\cite{Gutfleisch2011,Miyake2021}.
For a bcc FeCoNi ternary, the framework recovers the experimental Slater--Pauling peak, converging from three different starting compositions to Fe$_{0.79}$Co$_{0.21}$ ($\mu_0 M \approx 2.35$~T, where $\mu_0$ is the vacuum permeability).
Expanding the search to thirty candidate elements yields the same composition without enumerating candidates.
For LnT$_5$ rare-earth intermetallics (Ln = lanthanide, T = transition metal), the framework optimizes 43 composition parameters across three sublattices and identifies high-magnetization compositions including (Nd$_{0.983}$Pr$_{0.017}$)(Co$_{0.72}$Fe$_{0.28}$)$_2$Fe$_3$ and Pr(Co$_{0.80}$Fe$_{0.20}$)$_2$Fe$_3$.
Different starting compositions lead to distinct local optima, providing a range of promising candidates for experimental investigation.
To our knowledge, this is the first demonstration of gradient-based compositional optimization across dozens of candidate elements directly from first-principles calculations.

We then experimentally validate the framework by synthesizing five compositions obtained during optimization starting from SmCo$_5$ and measuring their magnetization.
These include the target composition (Sm$_{0.354}$Pr$_{0.325}$Nd$_{0.321}$)(Co$_{0.977}$Fe$_{0.017}$Mn$_{0.006}$)$_5$, containing six elements.
Across the five samples, the measurements confirm the overall increase in magnetization and agree with calculations for the measured compositions with a mean absolute relative deviation of 4.2\%.
These results demonstrate that the framework can guide experimentally validated dopant selection and composition tuning in multicomponent spaces that are difficult to explore exhaustively.

To demonstrate design through the DOS, we seek a half-metal, in which one spin channel is metallic and the other insulating.
Maximizing the spin polarization of the DOS around the Fermi level in a half-Heusler $XYZ$ compound yields FeMn(As$_{0.83}$P$_{0.17}$), close to the predicted half-metal FeMnAs~\cite{Feng2014}.
Together, these results establish a general framework for designing materials to meet diverse objectives, offering multiple candidates for experimental investigation.

\section{Results}\label{sec2}

\subsection{Rediscovery of the Slater--Pauling curve}\label{sec2.1}

To confirm that the framework works, we first maximize the magnetization $M$ of a ternary body-centered cubic (bcc) Fe$_{x_{\text{Fe}}}$Co$_{x_{\text{Co}}}$Ni$_{x_{\text{Ni}}}$ alloy, a problem whose optimum is already known.
Its magnetic moment takes its largest value when Co is added to Fe in the absence of Ni, known as the Slater--Pauling maximum~\cite{Slater1937,Pauling1938}.
This well-characterized optimum is the target the framework should recover.

\begin{figure}[!htbp]
\centering
\includegraphics[width=\textwidth]{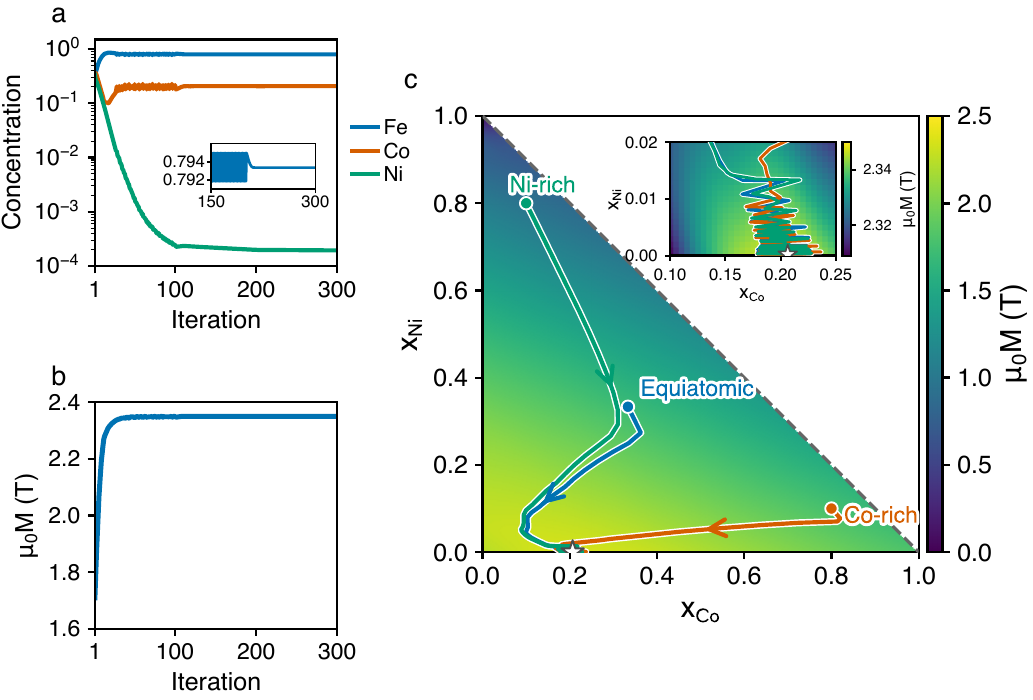}
\caption{\textbf{Optimization of bcc FeCoNi ternary alloy.} (a,b) A representative optimization starting from the equiatomic composition, showing (a) the composition and (b) the magnetization $\mu_0 M$ (in T) as functions of the iteration. The inset in (a) shows the Fe concentration over iterations 150--300. (c) The background color map shows $\mu_0 M$ computed across the ternary composition space by direct KKR-CPA evaluation. Each curve traces the gradient-ascent trajectory of a single optimization run projected onto the $x_{\text{Co}}$--$x_{\text{Ni}}$ plane. Three starting compositions---equiatomic (blue), Co-rich (vermillion), and Ni-rich (green)---all converge to $x_{\text{Fe}} \approx 0.79$, $x_{\text{Co}} \approx 0.21$, $x_{\text{Ni}} < 0.001$ (white star). The inset in (c) shows a finer map of the region around the maximum, computed with the same settings as the optimization. The trajectory endpoints coincide with the map maximum.}
\label{fig:feconi}
\end{figure}

First, we optimized the three composition ratios to maximize $M$, starting from the equiatomic composition $x_{\text{Fe}} = x_{\text{Co}} = x_{\text{Ni}} = 1/3$.
Figure~\ref{fig:feconi}a,b shows that, as the iterations proceed, the composition converges to $x_{\text{Fe}} \approx 0.79$, $x_{\text{Co}} \approx 0.21$, $x_{\text{Ni}} < 0.001$ (a), while $\mu_0 M$ rises monotonically to $\approx 2.35$~T (b).
To probe robustness of this result, we repeat the optimization from three widely separated compositions: equiatomic, Co-rich, and Ni-rich.
Figure~\ref{fig:feconi}c shows those trajectories in the ternary composition space over an independently computed KKR-CPA map of $\mu_0 M$.
All three converge to the same optimum independent of the starting point, and this optimum coincides with the maximum of that map.
This establishes the framework's reliability on a well-characterized landscape; the following sections apply it to higher-dimensional and multi-site design.
\FloatBarrier %
\subsection{Selection from thirty candidate elements}\label{sec2.2}

\begin{figure}[!htbp]
\centering
\includegraphics[width=\textwidth]{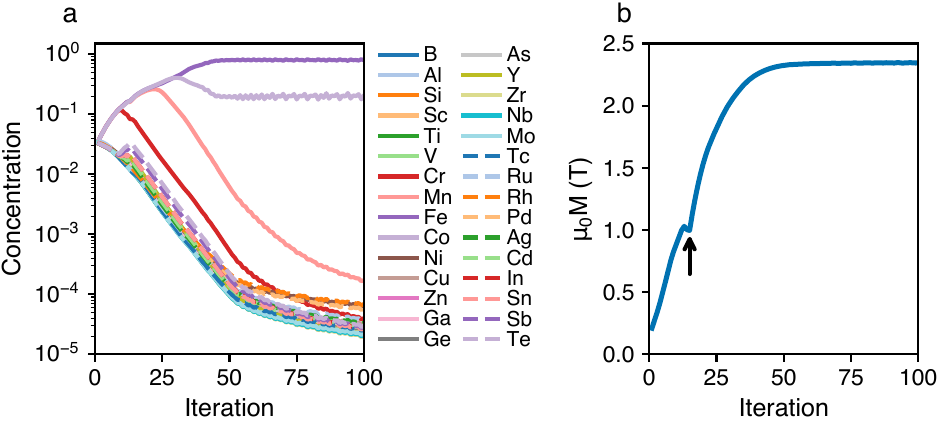}
\par\vspace{6pt}
\makebox[\textwidth][l]{\hspace{0.10\textwidth}{\sffamily\fontsize{9}{11}\selectfont c}}
{\fontsize{8}{9}\selectfont
\begin{tabular*}{0.82\textwidth}{@{}rl@{\extracolsep{\fill}}r@{}}
\hline
$N$ & Composition & $\mu_0 M$ (T) \\
\hline
93 & $\mathrm{Fe}_{0.79}\mathrm{Co}_{0.21}$ & 2.34 \\
2 & $\mathrm{Fe}_{0.87}\mathrm{Co}_{0.13}$ & 2.31 \\
1 & $\mathrm{Fe}$ & 2.28 \\
20 & $\mathrm{Mn}_{0.84}\mathrm{Te}_{0.16}$ & 2.01 \\
1 & $\mathrm{V}_{0.51}\mathrm{Te}_{0.49}$ & 0.61 \\
27 & (unconverged) & -- \\
\hline
\end{tabular*}
}
\caption{\textbf{Optimization of 30-element bcc alloy.} \textbf{a}, Composition evolution during magnetization optimization from an equiatomic initial composition. Only Fe and Co remain as major components at convergence. \textbf{b}, Magnetization $\mu_0 M$ vs optimization iteration, showing convergence after $\sim$50 iterations. The arrow marks a transient dip near iteration~15, reflecting a sign flip of the Cr local moment (antiferromagnetic alignment); the optimizer subsequently suppresses Cr and recovers monotonic convergence.
\textbf{c}, Representative compositions from 30-element alloy optimization (144 random initializations). 27 of the 144 runs did not satisfy the convergence criterion. $N$ gives the number of converged runs assigned to each group (Methods, Sec.~\ref{sec4.4}). $\mu_0 M$ is from re-evaluation with denser Brillouin-zone sampling.}
\label{fig:30elem}
\end{figure}

We next expand the candidate set to 30 elements and optimize the composition of the alloy $\mathrm{B}_{x_{1}}\allowbreak\mathrm{Al}_{x_{2}}\allowbreak\mathrm{Si}_{x_{3}}\allowbreak\mathrm{Sc}_{x_{4}}\allowbreak\mathrm{Ti}_{x_{5}}\allowbreak\mathrm{V}_{x_{6}}\allowbreak\mathrm{Cr}_{x_{7}}\allowbreak\mathrm{Mn}_{x_{8}}\allowbreak\mathrm{Fe}_{x_{9}}\allowbreak\mathrm{Co}_{x_{10}}\allowbreak\mathrm{Ni}_{x_{11}}\allowbreak\mathrm{Cu}_{x_{12}}\allowbreak\mathrm{Zn}_{x_{13}}\allowbreak\mathrm{Ga}_{x_{14}}\allowbreak\mathrm{Ge}_{x_{15}}\allowbreak\mathrm{As}_{x_{16}}\allowbreak\mathrm{Y}_{x_{17}}\allowbreak\mathrm{Zr}_{x_{18}}\allowbreak\mathrm{Nb}_{x_{19}}\allowbreak\mathrm{Mo}_{x_{20}}\allowbreak\mathrm{Tc}_{x_{21}}\allowbreak\mathrm{Ru}_{x_{22}}\allowbreak\mathrm{Rh}_{x_{23}}\allowbreak\mathrm{Pd}_{x_{24}}\allowbreak\mathrm{Ag}_{x_{25}}\allowbreak\mathrm{Cd}_{x_{26}}\allowbreak\mathrm{In}_{x_{27}}\allowbreak\mathrm{Sn}_{x_{28}}\allowbreak\mathrm{Sb}_{x_{29}}\allowbreak\mathrm{Te}_{x_{30}}$, where $x_i$ denotes the concentration of each element.
We use the same bcc lattice as in the FeCoNi case.
Gradient-based optimization addresses such many-element composition spaces directly, without enumerating compositions.
Fig.~\ref{fig:30elem}a shows the composition evolution starting from an equiatomic mixture of all 30 elements.
After approximately 50 iterations, only Fe (79\%) and Co (21\%) retain significant fractions, with the remaining 28 elements driven to negligible concentrations.
The optimizer thus selects the same high-moment Fe--Co composition as in the ternary case (Sec.~\ref{sec2.1}), from a 30-dimensional search space.
Figure~\ref{fig:30elem}b shows the magnetization rising steadily from the start as the composition is refined.
Near iteration~15, the curve briefly dips: at this point the local moment of Cr becomes aligned antiparallel to those of Fe and Co, so it subtracts from $M$ rather than adding to it.
The optimizer recovers by suppressing Cr and resumes its climb, showing that gradient-based optimization can navigate such non-smooth features in the landscape.
The per-iteration cost increased by only a factor of about 2.4 when moving from 3 to 30 compositional parameters (62~s vs 147~s).
This modest increase reflects additional scattering calculations for more elements in CPA, not an increase in gradient cost with the number of composition parameters.
The framework can thus scale to even larger compositional spaces without a proportional increase in computational effort.

We also ran the optimization from 144 random starting compositions and examined which optima they reached.
The converged runs yielded five representative compositions (Fig.~\ref{fig:30elem}c): the majority (93 runs) reached Fe$_{0.79}$Co$_{0.21}$, while 20 converged to Mn$_{0.84}$Te$_{0.16}$, two to Fe$_{0.87}$Co$_{0.13}$, and one each to Fe and V$_{0.51}$Te$_{0.49}$.
These results demonstrate that the framework can identify multiple distinct candidate materials by varying the initial composition.

\subsection{Multi-site design of permanent-magnet candidates}\label{sec2.3}

We next apply the framework to a multi-site permanent-magnet system.
The hexagonal LnT$_5$ system, which includes SmCo$_5$~\cite{Strnat1967} and its derivatives~\cite{Patrick2019,Daalderop1996,Okumura2023}, has three inequivalent crystallographic sites: Ln(1a), T(2c), and T(3g).
Each hosts a different set of candidate elements: 15 lanthanides (La through Lu) for Ln(1a), and 14 transition metals from the 3$d$ (Ti to Ni) and 4$d$ (Zr to Pd) series for each T site.
In total, this gives 43 parameters.
As in the other two demonstrations, the objective is the magnetization, $M$.
Note that the lanthanide 4$f$ moments are treated as fixed open-core contributions and enter the objective as a term linear in the concentrations (Methods, Sec.~\ref{sec4.4}).

\begin{figure}[!htbp]
\centering
\includegraphics[width=0.87\textwidth]{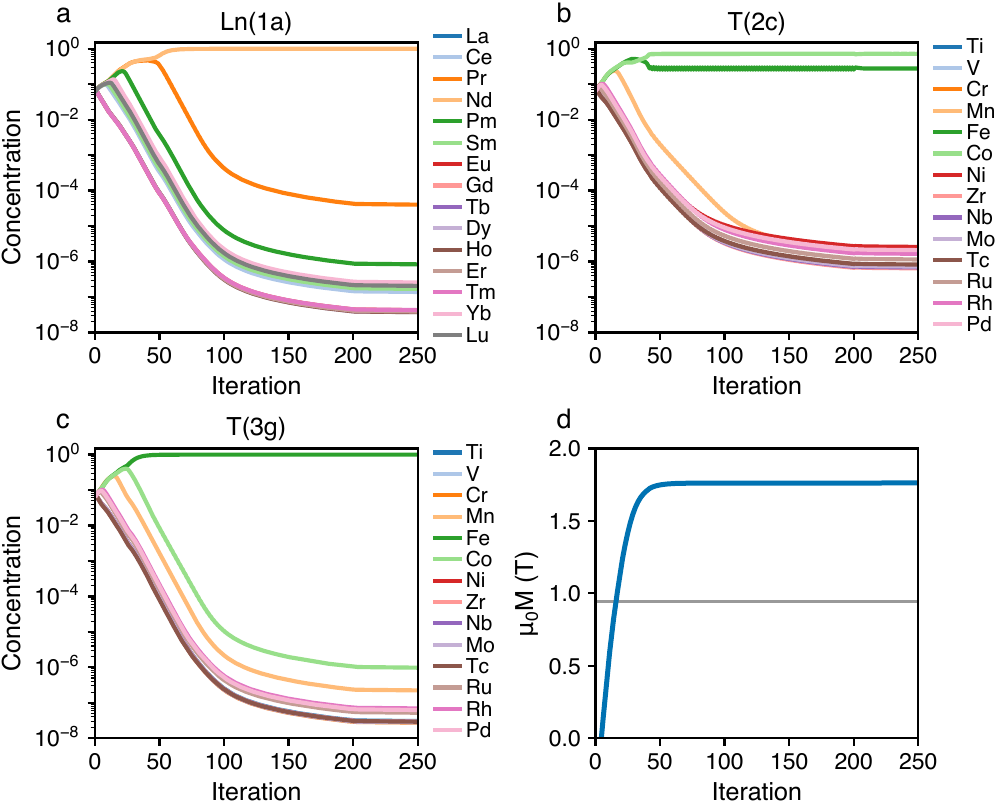}
\par\vspace{6pt}
\makebox[\textwidth][l]{\hspace{0.10\textwidth}{\sffamily\fontsize{9}{11}\selectfont e}}
{\fontsize{8}{9}\selectfont
\begin{tabular*}{0.82\textwidth}{@{}rl@{\extracolsep{\fill}}r@{}}
\hline
$N$ & Composition & $\mu_0 M$ (T) \\
\hline
-- & $\mathrm{Sm}\mathrm{Co}_{5}$ (reference) & 0.943 \\
\hline
8 & $\mathrm{Pr}(\mathrm{Co}_{0.799}\mathrm{Fe}_{0.201})_{2}\mathrm{Fe}_{3}$ & 1.757 \\
92 & $(\mathrm{Nd}_{0.983}\mathrm{Pr}_{0.017})(\mathrm{Co}_{0.719}\mathrm{Fe}_{0.281})_{2}\mathrm{Fe}_{3}$ & 1.746 \\
1 & $\mathrm{Pr}(\mathrm{Co}_{0.972}\mathrm{Mn}_{0.028})_{2}\mathrm{Fe}_{3}$ & 1.729 \\
2 & $\mathrm{Pr}(\mathrm{Fe}_{0.886}\mathrm{Co}_{0.114})_{2}\mathrm{Fe}_{3}$ & 1.674 \\
3 & $(\mathrm{Pr}_{0.913}\mathrm{Nd}_{0.087})(\mathrm{Fe}_{0.891}\mathrm{Co}_{0.109})_{2}\mathrm{Fe}_{3}$ & 1.672 \\
3 & $\mathrm{Pr}\mathrm{Fe}_{2}(\mathrm{Fe}_{0.981}\mathrm{Co}_{0.019})_{3}$ & 1.665 \\
1 & $\mathrm{Pr}(\mathrm{Mn}_{0.976}\mathrm{Fe}_{0.024})_{2}(\mathrm{Fe}_{0.959}\mathrm{Co}_{0.041})_{3}$ & 1.532 \\
2 & $\mathrm{Pr}\mathrm{Mn}_{2}(\mathrm{Ni}_{0.985}\mathrm{Co}_{0.015})_{3}$ & 1.461 \\
1 & $\mathrm{Pr}\mathrm{Mn}_{2}(\mathrm{Ni}_{0.766}\mathrm{Co}_{0.234})_{3}$ & 1.457 \\
2 & $\mathrm{Nd}(\mathrm{V}_{0.851}\mathrm{Ti}_{0.149})_{2}(\mathrm{V}_{0.966}\mathrm{Ti}_{0.034})_{3}$ & 0.493 \\
1 & $\mathrm{Pm}(\mathrm{V}_{0.936}\mathrm{Cr}_{0.064})_{2}(\mathrm{Ti}_{0.662}\mathrm{V}_{0.338})_{3}$ & 0.369 \\
28 & (unconverged) & -- \\
\hline
\end{tabular*}
}
\caption{\textbf{Optimization of LnT$_5$-type permanent-magnet system.} Composition evolution at each crystallographic site: \textbf{a}, Ln(1a) sublattice (15 lanthanide elements); \textbf{b}, T(2c) sublattice (14 transition metals); \textbf{c}, T(3g) sublattice (14 transition metals). Nd dominates the Ln site, while the T(2c) site converges to a Co--Fe mixture and the T(3g) site to nearly pure Fe. \textbf{d}, Magnetization $\mu_0 M$ including the 4$f$ open-core contribution, which is part of the objective. The grey line marks the SmCo$_5$ reference value.
\textbf{e}, Candidate LnT$_5$ compositions and their magnetization. Each representative composition was re-evaluated with progressively denser Brillouin-zone sampling. $N$ gives the number of optimization runs (out of 144 random starts) assigned to each group (Methods, Sec.~\ref{sec4.4}); the dash marks the SmCo$_5$ reference row, computed with the same method. $\mu_0 M$ includes the 4$f$ open-core correction.}
\label{fig:lnt5}
\end{figure}

Fig.~\ref{fig:lnt5}a--c shows the composition evolution during optimization, starting from an equiatomic mixture of the candidate elements at each site.
The optimization converges to $\mathrm{Nd}(\mathrm{Co}_{0.72}\mathrm{Fe}_{0.28})_2\mathrm{Fe}_3$: the Ln(1a) site is dominated by Nd, the T(2c) site converges to a Co--Fe mixture, and the T(3g) site to nearly pure Fe.
Its magnetization, $\mu_0 M \approx 1.75$~T, exceeds the 0.94~T of SmCo$_5$ (Fig.~\ref{fig:lnt5}d).
To our knowledge, this composition has not been experimentally reported.

We also run the optimization from 144 random starting compositions and obtain eleven representative compositions (Fig.~\ref{fig:lnt5}e).
The most frequently reached composition is $(\mathrm{Nd}_{0.983}\mathrm{Pr}_{0.017})(\mathrm{Co}_{0.72}\mathrm{Fe}_{0.28})_2\mathrm{Fe}_3$ (1.75~T, 92 runs), and $\mathrm{Pr}(\mathrm{Co}_{0.80}\mathrm{Fe}_{0.20})_2\mathrm{Fe}_3$ reaches the highest magnetization (1.757~T, 8 runs).
The selection of Nd and Pr is consistent with the established trend that replacing Sm with Pr or Nd increases the magnetization of LnCo$_5$~\cite{Buschow1977,Strnat1990}.
These results demonstrate that the framework can simultaneously optimize the compositions of multiple inequivalent crystallographic sites and identify candidates with higher calculated magnetization than SmCo$_5$.

\FloatBarrier %
\subsection{Doping optimization and experimental validation}\label{sec2.4}

In practical materials development, doping and elemental substitution in existing compounds are common approaches to improving their properties.
Here, we demonstrate that the framework can also guide such substitutions in SmCo$_5$.
After optimizing the dopant concentrations, we synthesize selected compositions and experimentally confirm the predicted increase in magnetization.

We select the candidate elements considering cost, availability, and handling risks.
We run two optimizations starting from SmCo$_5$ with different sets of substitution candidates for Sm.
One includes Y, La, Ce, Pr, Nd, Gd, Tb, Dy, and Ho, while the other includes Y, La, Ce, Pr, and Nd.
In both runs, the substitution candidates for Co are Ti, V, Mn, Fe, and Ni at both the 2c and 3g sites.
Sm and Co remain available at their respective sites throughout the optimization.
The objective is the magnetization, including the 4$f$ contribution (Methods, Sec.~\ref{sec4.4}).

\begin{figure}[!htbp]
\centering
\includegraphics[width=\textwidth]{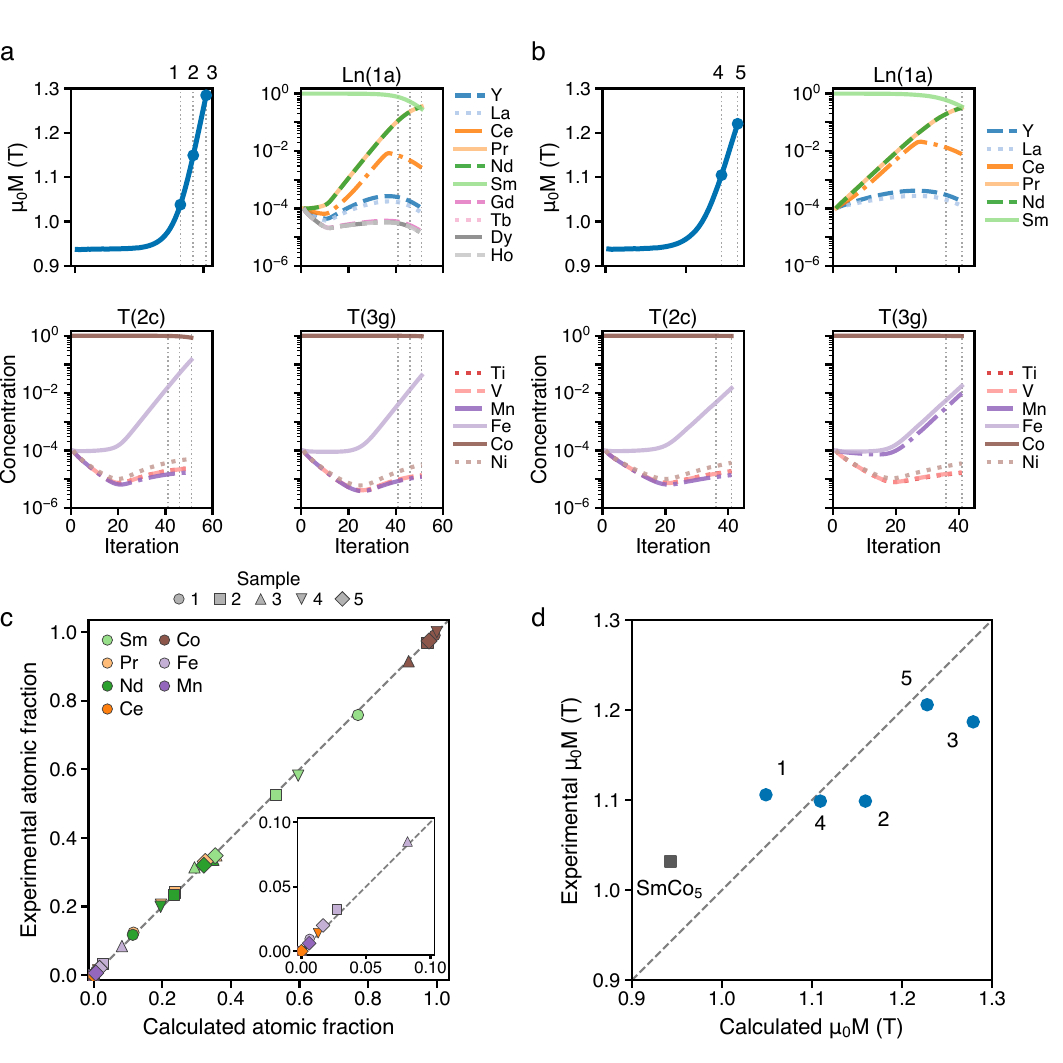}
\caption{\textbf{Magnetization of SmCo$_5$-based alloys.} \textbf{a,b}, Calculated saturation magnetization and compositions at Ln(1a), T(2c), and T(3g) along the trajectories from SmCo$_5$ with ten (\textbf{a}) and six (\textbf{b}) Ln-site candidates. Dotted lines mark the iterations selected for synthesis of Samples 1--3 (\textbf{a}) and 4 and 5 (\textbf{b}). \textbf{c}, Calculated atomic fractions compared with bulk ICP measurements, normalized separately for lanthanides and transition metals. \textbf{d}, Saturation magnetization $\mu_0 M$ measured at 300~K for the five samples, compared with the value recalculated with converged \textit{k}-point sampling for each measured composition. The square marks the SmCo$_5$ reference. The dashed lines in \textbf{c} and \textbf{d} mark equality. The 0~K calculation is compared directly with the 300~K measurement.}
\label{fig:experiment}
\end{figure}

Both trajectories replace Sm with Pr and Nd and Co with Fe, moving toward PrFe$_5$ (Fig.~\ref{fig:experiment}a,b).
We select three intermediate compositions from the first trajectory (Samples 1--3; Fig.~\ref{fig:experiment}a) and two from the second (Samples 4 and 5; Fig.~\ref{fig:experiment}b), at the iterations marked by dotted lines.
Examples include (Sm$_{0.770}$Pr$_{0.116}$Nd$_{0.114}$)(Co$_{0.994}$Fe$_{0.006}$)$_5$ for Sample 1 and (Sm$_{0.354}$Pr$_{0.325}$Nd$_{0.321}$)(Co$_{0.977}$Fe$_{0.017}$Mn$_{0.006}$)$_5$ for Sample 5; all five target compositions are listed in Extended Data Table~\ref{tab:experiment_compositions}.

To test these predictions experimentally, we synthesize five samples using the selected compositions as targets.
The measured compositions closely match the targets (Fig.~\ref{fig:experiment}c; Extended Data Table~\ref{tab:experiment_compositions}), and all five samples form the CaCu$_5$-type phase as the main phase (Extended Data Table~\ref{tab:experiment_properties}).
We then measure their saturation magnetization $\mu_0 M$ at 300~K.

To compare the predictions with the measurements, we recalculate the magnetization using the measured composition of each sample.
Figure~\ref{fig:experiment}d compares these calculated and measured values, together with those for the SmCo$_5$ reference.
All five samples show higher measured magnetization than the reference.
For these five samples, the correlation coefficient between the calculated and measured values is 0.83, and the mean absolute relative deviation is 4.2\%.

The remaining differences may reflect secondary phases, finite-temperature effects, the assumed lattice parameters, and approximations in the electronic-structure calculations.
These results demonstrate that the framework can guide dopant selection and quantitatively predict the magnetization of the resulting materials.

\FloatBarrier %
\subsection{Half-metal design using a density-of-states objective}\label{sec2.5}

\begin{figure}[!htbp]
\centering
\includegraphics[width=\textwidth]{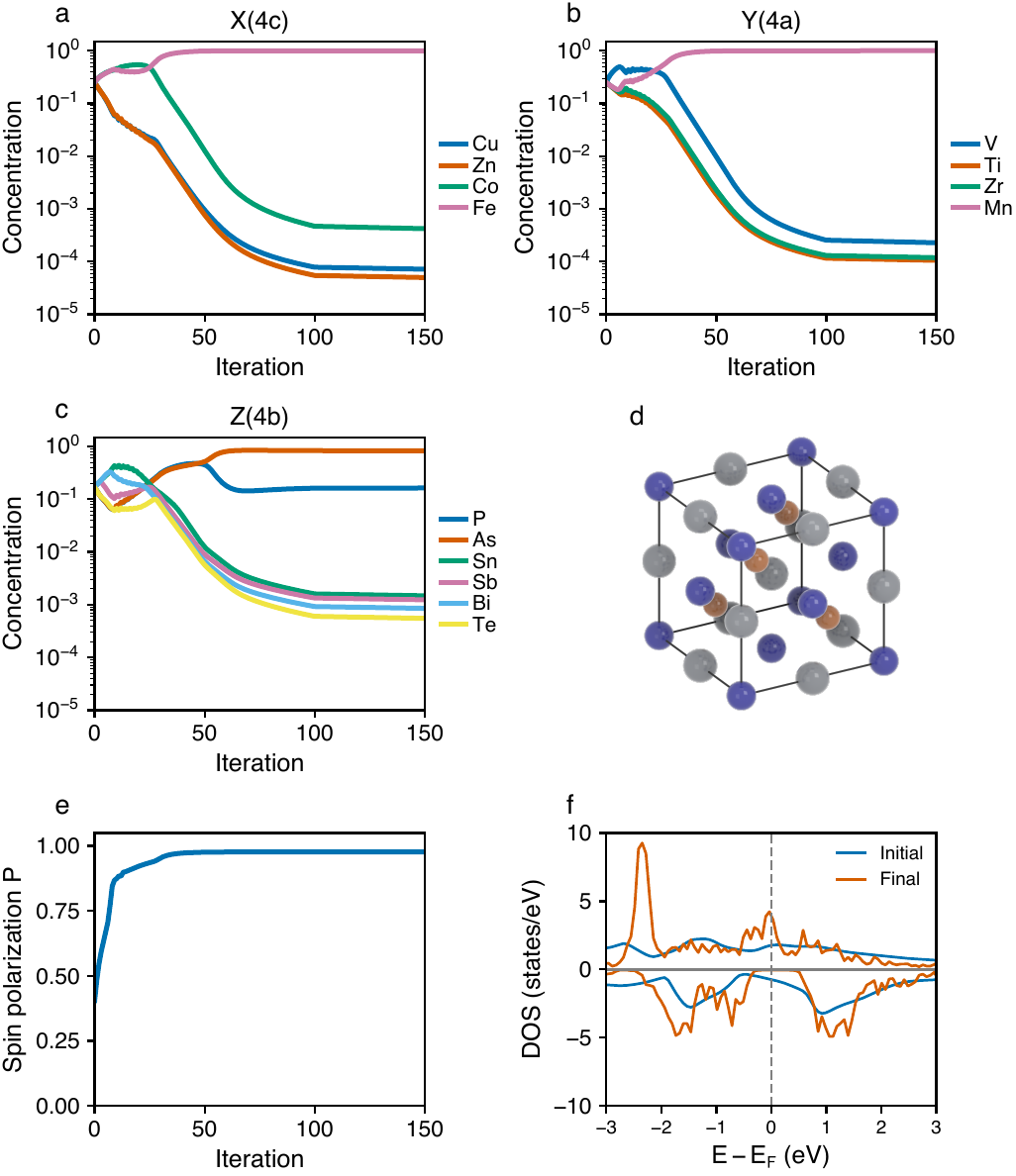}
\caption{\textbf{Gradient-based half-metal design in a 14-parameter half-Heusler space.} All three inequivalent crystallographic sites of the $XYZ$ half-Heusler structure are optimized simultaneously using the spin polarization $P$ of the window-averaged DOS at the Fermi level (Eq.~\ref{eq:halfmetal}). \textbf{a}--\textbf{c}, Composition evolution at the $X$, $Y$, and $Z$ sites (4, 4, and 6 candidate elements, respectively). \textbf{d}, Half-Heusler crystal structure (space group $\mathrm{F}\bar{4}3\mathrm{m}$). \textbf{e}, Spin polarization $P$ versus iteration. \textbf{f}, Initial (equiatomic) and final spin-resolved DOS. The dashed vertical line marks the Fermi level.}
\label{fig:heusler}
\end{figure}

So far, we have optimized magnetization, a scalar physical quantity.
We next show that the framework can also target specific features of the electronic structure by defining an objective from the density of states (DOS).
As an example, we seek a half-metal, in which the DOS at the Fermi level vanishes in one spin channel while remaining finite in the other.
To find such a material, we maximize the spin polarization of the DOS around the Fermi level, defined as
\begin{equation}
P = \frac{\langle n_{\mathrm{maj}}\rangle_\delta - \langle n_{\mathrm{min}}\rangle_\delta}{\langle n_{\mathrm{maj}}\rangle_\delta + \langle n_{\mathrm{min}}\rangle_\delta}.
\label{eq:halfmetal}
\end{equation}
Here, $\langle n_\sigma\rangle_\delta$ denotes the spin-resolved DOS averaged over an energy window of half-width $\delta$ centred at the Fermi level:
\begin{equation*}
\langle n_\sigma\rangle_\delta = \frac{1}{2\delta}\int_{E_{\mathrm{F}}-\delta}^{E_{\mathrm{F}}+\delta} n_\sigma(E)\,dE.
\end{equation*}
The quantity $n_{\mathrm{maj}}$ ($n_{\mathrm{min}}$) denotes the DOS of the majority (minority) spin channel around the Fermi level.
This definition ensures that $P$ remains unchanged when the spin labels are exchanged.
The window averaging helps stabilize the optimization by reducing its sensitivity to fine features of the DOS around $E_{\mathrm{F}}$.
We set $\delta = 0.010$~Ry $\approx 0.136$~eV.
For reference, NiMnSb has a calculated minority-spin gap of about 0.5~eV~\cite{Galanakis2002}.
Our full averaging width, $2\delta \approx 0.272$~eV, is chosen to be smaller than this gap.
The upper bound $P = 1$ corresponds to an ideal half-metal in which the minority window-average vanishes.
We consider the half-Heusler $XYZ$ structure (space group $\mathrm{F}\bar{4}3\mathrm{m}$, Fig.~\ref{fig:heusler}d), a canonical family of predicted half-metals~\cite{deGroot1983}.
We placed composition variables on the three inequivalent crystallographic sites simultaneously: 4 candidates on the $X$ site (Cu, Zn, Co, Fe), 4 on the $Y$ site (V, Ti, Zr, Mn), and 6 on the $Z$ site ($sp$ elements P, As, Sn, Sb, Bi, Te), giving 14 parameters in total.

Figure~\ref{fig:heusler}a--c shows the changes in composition at each site during optimization.
The optimization started from an equiatomic composition on all three sites.
As the optimization progressed, the $X$ and $Y$ sites became almost entirely occupied by Fe and Mn, respectively, while the $Z$ site became predominantly As with a small amount of P.
The final composition was FeMn(As$_{0.83}$P$_{0.17}$).
This is close to FeMnAs, which has been predicted to be a half-metal by density functional theory~\cite{Feng2014}.
Figure~\ref{fig:heusler}e shows $P$ rising toward the bound, from 0.41 at the equiatomic start to 0.98.
The final DOS (Fig.~\ref{fig:heusler}f) shows clear half-metallic behaviour: the minority-spin DOS is suppressed over an approximately 0.6~eV window around $E_{\mathrm{F}}$, while the majority channel remains metallic.
The small deviation from $P = 1$ is attributed to numerical broadening.

\FloatBarrier %
\section{Discussion}\label{sec3}

KKR-CPA allows composition to be treated as a continuous variable representing the proportions of the constituent elements.
By integrating reverse-mode AD~\cite{Griewank2008,Baydin2018}, we can compute composition gradients of target properties.
This combination provides four key capabilities.
First, composition can be optimized efficiently by updating the elemental concentrations using the calculated gradients.
Second, every composition explored during optimization is evaluated directly by first-principles calculations, so the predicted properties do not rely on a trained surrogate model.
Third, the search can include many candidate elements simultaneously, since all composition gradients are computed in a single backward pass.
Fourth, any differentiable quantity computable within KKR-CPA can serve as an optimization objective.
Our results demonstrate these four capabilities in practice: we performed gradient-based optimization of up to 43 composition parameters, evaluating every composition directly by first-principles calculations.
The same framework was applied to two distinct objectives: magnetization and spin polarization of the DOS near the Fermi level.
Using these capabilities, the framework recovered the known Fe--Co optimum and identified new candidate compositions for both permanent magnets and half-metals.
Experiments on five SmCo$_5$-based samples further confirmed the predicted enhancement in magnetization, with measured values agreeing closely with the calculations.
Together, these results demonstrate that the framework can guide the discovery of new materials from a wide range of candidate elements, with improvements in their properties verified experimentally.

No existing method offers all four at once.
Combining conventional plane-wave DFT~\cite{Schmitz2026} or tight-binding~\cite{Friede2024} with AD, for example, has not been demonstrated for materials design by composition in these studies.
Effective-atom theory~\cite{Tahmassebpur2025} is limited to objectives that are functions of the energy, and only its final, discrete compositions are physical.
Our framework therefore offers a new approach to materials design. It builds materials from scratch, pinpointing entirely unknown compositions with the desired properties. It also simultaneously optimizes multiple dopant concentrations in existing materials without an exhaustive search. This application may have even greater practical value.

Our framework can optimize differentiable targets accessible to KKR-CPA, including user-specified band structures~\cite{Faulkner1980}, magnetic order~\cite{Liechtenstein1987}, Curie temperatures~\cite{Gyorffy1985,Takahashi2007}, magnetic anisotropy~\cite{Okumura2023}, and transport properties~\cite{Lowitzer2010}.
Each becomes an optimization target simply by adding the routine that computes it.
The objective need not be a single physical property: multiple targets can be combined, and even quantities unrelated to the electronic structure can be included.
For example, adding the concentration-weighted price to the objective leads to lower-cost materials.

Finally, we turn to the current limitations of the framework and the future directions they suggest.
Although the optimization recovers the maximum of the calculated magnetization landscape for bcc FeCoNi, the predicted composition differs slightly from the experimental Slater--Pauling peak.
Agreement with experiment also depends on the accuracy of the underlying KKR-CPA calculations.
The present implementation uses the atomic-sphere approximation; a full-potential treatment would improve the accuracy of the underlying electronic-structure calculation.
Thermodynamic stability and lattice relaxation are not yet addressed, so the outputs are compositions on a fixed lattice rather than relaxed equilibrium phases.
However, because the total energy can be differentiated with respect to the atomic positions, lattice relaxation can be brought into the same framework, lifting this limitation in future work.
Differentiation can also be extended to external parameters such as applied magnetic and electric fields.
Together, these extensions would enable the joint inverse design of composition, lattice, and external fields to achieve a desired property.
This would make first-principles calculations a general tool for materials design, offering a direct and physically grounded route from target properties to materials that realize them.
\section{Methods}\label{sec4}

\subsection{KKR-CPA Method}\label{sec4.2}

The KKR method~\cite{Korringa1947,Kohn1954,Ebert2011KKR,Faulkner2018} solves the Kohn--Sham equations of density functional theory~\cite{Hohenberg1964,Kohn1965} via multiple scattering theory, computing the single-particle Green's function directly rather than solving an eigenvalue problem for Bloch eigenstates.
In multiple scattering theory, the scattering of electrons from each atomic potential is described by a $t$-matrix, the operator that maps an incident wave to the scattered wave at energy $E$.
Within the atomic-sphere approximation (ASA)~\cite{Andersen1975} adopted here, the atomic potentials are spherically symmetric, and the $t$-matrix is diagonal in the angular-momentum quantum numbers $L = (l, m)$, where $l$ is the orbital and $m$ the magnetic quantum number.
It is then characterized by the radial elements $t_l^{(s)}(E)$ at each inequivalent site $s$ in the basis.
These elements are obtained by solving the radial Schr\"odinger equation within the atomic sphere.
The resulting phase shifts $\eta_l^{(s)}(E)$, defined by the asymptotic phase of the partial-wave radial wavefunction at large distance, determine $t_l^{(s)}$.
The propagation of scattered waves between sites is described by the structure-constant matrix $g_{LL'}^{ss'}(\mathbf{k}, E)$, which depends on the crystal geometry and energy but not on the atomic potentials.
The central quantity is the scattering path operator
\begin{equation}
\tau(\mathbf{k}, E) = \left[t^{-1}(E) - g(\mathbf{k}, E)\right]^{-1},
\end{equation}
which sums all multiple-scattering events in the crystal.
Here $\tau$, $t$, and $g$ are matrices in the joint $(s, L)$ space; $t^{-1}(E)$ is block-diagonal over inequivalent sites with on-site elements $\delta_{LL'}/t_l^{(s)}(E)$, and the matrix inversion runs over the joint $(s, L)$ space.
Physical observables are obtained from the site-diagonal Green's function $G_s(\mathbf{r}, \mathbf{r}'; E)$, constructed from $\tau$ by partial-wave decomposition with the regular and irregular radial solutions of the Schr\"odinger equation within each atomic sphere~\cite{Faulkner1980,Ebert2011KKR}.
The central matrix entering this construction is the on-site BZ-integrated scattering path operator $T_{LL'}^{(s)}(E) = \frac{1}{\Omega_{\mathrm{BZ}}}\int_{\mathrm{BZ}} d^{3}\mathbf{k}\,\tau_{LL'}^{ss}(\mathbf{k}, E)$.

The CPA~\cite{Soven1967,Velicky1968,Gyorffy1972,Akai1989} describes the electronic structure of substitutionally disordered alloys by replacing the random distribution of atomic potentials with a self-consistent effective medium.
Each element occupies site $s$ with a probability given by its concentration $x_i^{(s)}$, so the composition is described by continuous variables.
This formulation allows the objective function $\mathcal{O}(\mathbf{x})$ to be differentiated with respect to composition.
At each inequivalent site $s$, this effective medium is represented by a coherent $t$-matrix $\tilde{t}^{(s)}(E)$.
Define the impurity-embedded scattering path operator
\begin{equation}
\tau_i^{(s)}(E) = \left[(t_i^{(s)})^{-1} - (\tilde{t}^{(s)})^{-1} + (T^{(s)})^{-1}\right]^{-1},
\end{equation}
where $t_i^{(s)}(E)$ is the single-site $t$-matrix of element $i$ at site $s$.
The coherent $t$-matrix is determined by the CPA self-consistency condition
\begin{equation}
\sum_i x_i^{(s)} \tau_i^{(s)}(E) = T^{(s)}(E),
\end{equation}
which holds independently on each inequivalent site $s$ and defines $\tilde{t}^{(s)}$ implicitly as a function of $\mathbf{x}$.
Because the condition is smooth in $\tilde{t}^{(s)}$ and enters $x_i^{(s)}$ linearly, the implicit function theorem ensures that $\tilde{t}^{(s)}$ depends smoothly on $\mathbf{x}$; the Green's function and any observable derived from it inherit this smoothness.
In particular, the spin-resolved charge density is
\begin{equation}
\rho_\sigma(\mathbf{r};\, \mathbf{x}) = -\frac{1}{\pi}\Im\int_{-\infty}^{E_{\mathrm{F}}} dE\, G_\sigma(\mathbf{r}, \mathbf{r};\, E;\, \mathbf{x}),
\end{equation}
where $\sigma$ labels the spin channel.
The two observables optimized in this work follow directly from these quantities.
The total magnetic moment is the integrated spin density,
\begin{equation}
\mu_{\mathrm{tot}}(\mathbf{x}) = \int \left[\rho_\uparrow(\mathbf{r};\, \mathbf{x}) - \rho_\downarrow(\mathbf{r};\, \mathbf{x})\right] d\mathbf{r},
\end{equation}
and the spin-resolved density of states is
\begin{equation}
n_\sigma(E;\, \mathbf{x}) = -\frac{1}{\pi}\Im \int G_\sigma(\mathbf{r}, \mathbf{r};\, E;\, \mathbf{x})\, d\mathbf{r},
\end{equation}
with both integrals taken over the unit cell.
Our calculations are scalar-relativistic and use the spin-polarized local-density exchange-correlation functional in the Moruzzi--Janak--Williams parametrization~\cite{Moruzzi1978}.
Our KKR-CPA implementation is based on the AkaiKKR software~\cite{Akai1989}.

\subsection{Automatic Differentiation}\label{sec4.3}

Automatic differentiation (AD) is a method for computing derivatives of functions implemented as computer programs~\cite{Baydin2018,Griewank2008}.
Computational simulations, including DFT calculations, can be expressed as sequences of numerical operations, such as arithmetic and matrix operations.
When these operations are differentiable, the derivative of the overall calculation can be obtained by combining their local derivatives through the chain rule.
The dependencies between the operations form a computational graph.

AD has two modes that differ in how they traverse the computational graph.
Forward mode carries derivatives from the inputs toward the output as the function is evaluated; a full gradient then costs one pass per input parameter, so forward mode is efficient when the outputs outnumber the inputs.
Reverse mode instead carries derivatives from the output back to the inputs, and is efficient in the opposite regime, in which many inputs feed a few outputs.
Inverse materials design is precisely this regime---many composition parameters and a single scalar objective---so we use reverse mode.

Reverse-mode AD consists of a forward pass and a backward pass.
Formally, an evaluation of the objective is a composition of elementary operations,
\begin{equation}
\mathcal{O}(\mathbf{x}) = (f_L \circ f_{L-1} \circ \cdots \circ f_1)(\mathbf{x}),
\end{equation}
with intermediate variables $\mathbf{v}_k = f_k(\mathbf{v}_{k-1})$, $\mathbf{v}_0 = \mathbf{x}$, and $\mathbf{v}_L = \mathcal{O}$.
The forward pass evaluates the objective and saves intermediate information needed for differentiation, which may include both the values $\mathbf{v}_k$ and local derivative coefficients.
Using the known derivative rules of the elementary operations and the saved information, the backward pass propagates the adjoint variables $\bar{\mathbf{v}}_k \equiv \nabla_{\mathbf{v}_k} \mathcal{O}$ from the output back to the inputs through the chain rule,
\begin{equation}
\bar{\mathbf{v}}_{k-1} = \left(\frac{\partial f_k}{\partial \mathbf{v}_{k-1}}\right)^{\!\top} \bar{\mathbf{v}}_k,
\qquad \bar{\mathbf{v}}_L = 1,
\end{equation}
and a single sweep terminates with the full gradient $\nabla_{\mathbf{x}}\mathcal{O} = \bar{\mathbf{v}}_0$.
A single backward pass computes the derivatives of the objective with respect to all input parameters.
Its computational cost is typically only a few times that of evaluating the objective, without requiring a separate calculation for each parameter.
This makes reverse-mode AD particularly useful for optimizing materials with many composition parameters.
However, retaining intermediate values for the backward pass can require substantial memory.

Reverse-mode AD of this kind is now provided by mature software libraries such as PyTorch~\cite{Paszke2019}, Zygote~\cite{Innes2019}, and JAX~\cite{Bradbury2018}; we implement our differentiable KKR-CPA (AkaiKKR) in Python using JAX.

\subsection{Gradient-Based Inverse Materials Design Framework}\label{sec:framework}

We build the framework by re-implementing the KKR-CPA method of AkaiKKR~\cite{Akai1989} in JAX~\cite{Bradbury2018}, which makes the calculation automatically differentiable.
Reverse-mode AD then returns the gradient of any objective $\mathcal{O}$ with respect to the composition, $\nabla_{\mathbf{x}}\mathcal{O}$, which drives the gradient-based inverse-design loop of Fig.~\ref{fig:overview}.

Several implementation choices make this practical.
To take advantage of the fast AkaiKKR implementation, the gradient is computed in two stages that combine the original Fortran code with a JAX re-implementation.
The Fortran code first iterates the SCF cycle to convergence and exports the converged radial potentials, the CPA coherent $t$-matrix, and the energy and Brillouin-zone meshes.
A JAX re-implementation then rebuilds the single-site phase shifts and the CPA Green's function from this fixed-point state in a single pass and applies reverse-mode AD to the composition to obtain the explicit contribution to $\nabla_{\mathbf{x}}\mathcal{O}$.
This avoids differentiating through the SCF loop itself: at the converged fixed point, the implicit function theorem guarantees that total derivatives of observables with respect to composition are well defined without unrolling the iterative history~\cite{Christianson1998,Bai2019,Maliyov2025}.
The JAX evaluator reuses the converged self-consistent potentials and the geometry-dependent structure constants from the Fortran fixed point, holding them fixed and not backpropagating through them.
We then include the response of the self-consistent state by solving an adjoint linear equation derived from the self-consistency condition~\cite{Christianson1998,Blondel2022,Bai2019,Maliyov2025}.
A comparison with finite differences confirms that the difference is small in practice.

Here, we treat the composition ratios as the parameters to be optimized.
The composition for inequivalent sites is defined as $\mathbf{x} = (\mathbf{x}^{(1)}, \ldots, \mathbf{x}^{(S)})$, where $S$ is the number of inequivalent sites, and each site carries $\mathbf{x}^{(s)} = (x_1^{(s)}, \ldots, x_{n_s}^{(s)})$ with $\sum_i x_i^{(s)} = 1$ and $x_i^{(s)} \ge 0$.
Here, $i$ is the element index and $s$ is the site index.
To satisfy the constraints, we reparametrize it by a softmax,
\begin{equation}
x_i^{(s)} = \frac{\exp(\theta_i^{(s)})}{\sum_j \exp(\theta_j^{(s)})},
\label{eq:softmax}
\end{equation}
with unconstrained parameters $\boldsymbol{\theta}^{(s)}$, and update them using the gradient.
As a simple example, gradient descent (ascent) gives
\begin{equation}
\boldsymbol{\theta}^{[k+1]} = \boldsymbol{\theta}^{[k]} \mp \alpha_k \, \nabla_{\boldsymbol{\theta}} \mathcal{O}(\mathbf{x}(\boldsymbol{\theta}^{[k]})),
\end{equation}
where $\alpha_k > 0$ is the step size and $-$ ($+$) applies for minimization (maximization).

\subsection{Computational Details}\label{sec4.4}

\noindent \textbf{KKR-CPA parameters:} The partial-wave expansion was truncated at $l_{\text{max}} = 2$ for FeCoNi and 30-element bcc systems and for the half-Heusler system and $l_{\text{max}} = 3$ for LnT$_5$ systems. Lanthanide 4$f$ electrons were treated in open core, which is sufficient because the localized 4$f$ moment is essentially atomic and enters only as a fixed contribution per lanthanide atom. Brillouin-zone integration used a uniform $k$-point mesh, reduced to the irreducible zone by point-group and time-reversal symmetries. Mesh sizes for optimization were $20 \times 20 \times 20$ for the bcc systems, $12 \times 12 \times 12$ for the half-Heusler system, and $6 \times 6 \times 8$ for the hexagonal LnT$_5$ systems. For the spin-polarization objective, the DOS was evaluated on 200 energy points across the window.
For the final half-Heusler DOS, the finite imaginary energy shift $\eta = 0.001$~Ry introduces Lorentzian broadening.

\noindent \textbf{Self-consistency:} The main SCF loop used Broyden mixing with an initial mixing parameter $p_{\mathrm{mix}} = 0.01$ and up to 300 iterations. Convergence of each SCF run was judged from the stability of the total energy and charge neutrality over the final iterations, together with the residual based on the root-mean-square difference between the potential used to calculate the electron density and the potential obtained from that density. For calculations that did not converge, we changed the mixing parameters and retried the SCF calculation as follows. We first reduced $p_{\mathrm{mix}}$ to $0.005$ and then $0.001$. If convergence was still not achieved, we switched to Chebyshev acceleration with $p_{\mathrm{mix}} = 0.001$. Each retry allowed up to 300 iterations. To improve SCF convergence, we also varied the energy range of the integration contour.

\noindent \textbf{Optimization parameters:} We used the RMSprop optimizer~\cite{Bottou2018} provided by \texttt{optax}~\cite{Hessel2020}. RMSprop was run with decay rate 0.5 and numerical stability constant $10^{-8}$. A piecewise-constant learning-rate schedule was used with initial learning rate $0.1$, reduced by a factor of 10 at iterations 100, 200 (FeCoNi, 30-element, and half-Heusler) and 200, 300 (LnT$_5$). For the two doping optimizations described in Sec.~\ref{sec2.4}, we kept the learning rate fixed at 0.1 and evaluated the composition gradient using only the explicit contribution, without the response term described in Sec.~\ref{sec:framework}.

\noindent \textbf{Convergence and verification:} We assessed convergence at the final composition using the Frank--Wolfe gap~\cite{LacosteJulien2016}. A run was considered converged only if this gap fell below the corresponding empirically chosen threshold: $4 \times 10^{-3}$~T for LnT$_5$ and 0.058~T for the 30-element alloy. For converged runs, concentrations $x_i^{(s)} < 0.01$ were discarded and the remaining concentrations renormalized. Runs were grouped by comparing all pairs of final compositions, with a maximum difference of 0.05 in each elemental concentration at each site. Each group was represented by the run with the largest objective value. For each representative composition, convergence of the objective value was checked using progressively denser $k$-point meshes.

\noindent \textbf{Crystal structures:} FeCoNi and 30-element alloys were modeled with the bcc structure. LnT$_5$ systems were modeled with the hexagonal CaCu$_5$-type structure ($a = 9.388$ Bohr, $c/a = 0.798$). For the bcc systems, lattice parameters were updated at each optimization step by a Vegard-like rule applied to atomic volumes: the lattice constant was computed from the composition-weighted average of the elements' experimental atomic volumes. The gradient calculation also accounts for changes in the lattice constant with composition. The objective is the magnetization $M = \mu_{\mathrm{tot}}/V$ with $V$ the volume per atom of this lattice.

\noindent \textbf{Treatment of lanthanide 4$f$ moments:} For LnT$_5$, we used the open-core approximation and assigned a fixed 4$f$ magnetic moment to each lanthanide element. To calculate the total magnetic moment $\mu_{\mathrm{tot}}$, we weighted these fixed 4$f$ moments by the corresponding concentrations and added their sum to the contribution from the valence electrons. The 4$f$ moment per lanthanide atom was taken as the saturated Hund's-rule value with the 4$f$ spin antiparallel to the transition-metal moment: $+3~\mu_{\mathrm{B}}$ for Pr and Nd, $+2$ for Pm, $0$ for La, Ce, Sm, Yb, and Lu, $-7$ for Eu, Gd, and Tm, $-9$ for Tb and Er, and $-10$ for Dy and Ho.

\noindent \textbf{Computational time:} Calculations were run on a system with AMD EPYC 7702 processors. Median per-iteration wall-clock times were: the three FeCoNi runs, 62--65~s; the 30-element equiatomic run, 147~s; the 30-element random-start runs, 171~s; the LnT$_5$ equiatomic run, 23~min; the LnT$_5$ random-start runs, 9.8~min; and the half-metal, 21~min.

\subsection{Experimental Details}\label{sec4.5}

Samples were prepared and characterized at Toyota Motor Corporation.
Raw metals were weighed to a nominal rare-earth to transition-metal atomic ratio of 1:5.4.
The metals were alloyed by high-frequency induction melting and homogenized at 1000~$^\circ$C for 24~h.
The ingots were crushed and sieved to powders finer than 20~$\mu$m.
The composition of each powder was determined by inductively coupled plasma (ICP) analysis.
Lattice parameters were determined by X-ray diffraction, and the density by pycnometry.
The main-phase volume fraction was estimated from the ICP atomic ratios assuming an LnT$_5$ main phase and an Ln$_2$T$_{17}$ secondary phase with fixed densities of 8.25 and 7.82~g/cm$^3$, respectively.
The powders were fixed in epoxy resin and aligned by applying a static field of 1~T during curing.
Magnetization curves of the powders were measured using a Physical Property Measurement System (PPMS) between 300 and 453~K, with a maximum applied field of 7~T.
The saturation magnetization was determined by fitting the easy-axis magnetization curves over 4000--5570~kA/m using the law of approach to saturation.
The values reported here are those at 300~K, converted using the measured densities, and are not corrected for secondary phases.
The compositions in Extended Data Table~\ref{tab:experiment_compositions} are normalized within the lanthanide and transition-metal groups.

\backmatter

\bmhead{Acknowledgements}

This work was financially supported by Toyota Motor Corporation.
This work was supported by MEXT Next-Generation Computational Science Global Reach \& Innovation Program (Japan Grant Number JPMXP1920260216) and used computational resources of supercomputer Fugaku provided by the RIKEN Center for Computational Science (Project ID: hp260509).
The authors thank the Supercomputer Center, the Institute for Solid State Physics (ISSP), the University of Tokyo for the use of the facilities (ISSP-kyodo-SC-2025-D-0006, 2026-Ea-0016).
T.F. and H.S. acknowledge the Earth Simulator Center of the Japan Agency for Marine-Earth Science and Technology (JAMSTEC) for providing computational resources.
We thank Y. Konishi and M. Noda of Academeia Co., Ltd.\ for their help in rewriting the Fortran code into Python using JAX.
We also thank T. Ozaki, N. Kawashima, T. Misawa, K. Yoshimi, and K. Ido (ISSP); and T. Miyake (National Institute of Advanced Industrial Science and Technology); and M. Saito and T. Shoji (Toyota Motor Corporation) for helpful discussions.
Claude (Anthropic) and Codex (OpenAI) assisted with drafting and editing the manuscript.

\section*{Declarations}

\noindent \textbf{Competing interests}: H.A. is affiliated with Academeia Co., Ltd. K.I. (Inui) is affiliated with Calmarion Inc. The other authors declare no competing interests.

\noindent \textbf{Data availability}: The processed data (optimization trajectories, converged compositions, and derived magnetic properties) used to produce the figures and tables, and the raw Fortran KKR-CPA outputs, are available from the corresponding authors upon reasonable request.

\noindent \textbf{Code availability}: The base KKR-CPA code, AkaiKKR (Machikaneyama)~\cite{Akai1989}, is available at \url{http://kkr.issp.u-tokyo.ac.jp/} (registration required). The differentiable JAX implementation and optimization code developed for this work are available from the corresponding authors upon reasonable request.

\noindent \textbf{Author contributions}: K.I. (Inui) conceived of the study, designed the framework, and supervised the work. K.I. (Ishii) developed the framework and performed the calculations. H.A. and T.F. contributed methodology. H.S. provided the reference data. H.H. and N.S. contributed to sample synthesis and magnetic characterization.
K.I. (Ishii) and K.I. (Inui) wrote the manuscript; all authors reviewed and revised it.

\clearpage
\section*{Extended Data}

\setcounter{table}{0}
\renewcommand{\theHtable}{ED.\arabic{table}}
\renewcommand{\tablename}{Extended Data Table}

\begin{table}[htbp]
\centering
\small
\caption{\textbf{Target and measured sample compositions.} Target compositions are those proposed for synthesis. Measured compositions were obtained by inductively coupled plasma (ICP) analysis of the bulk samples. Atomic fractions are normalized separately for lanthanides and transition metals. Group totals may differ from unity because of rounding.}
\label{tab:experiment_compositions}
\begin{tabular}{@{}llrrrrrrr@{}}
\hline
Sample & & \multicolumn{4}{c}{Lanthanides} & \multicolumn{3}{c}{Transition metals} \\
 & & Sm & Pr & Nd & Ce & Co & Fe & Mn \\
\hline
\multirow{2}{*}{Sample 1} & Target & 0.770 & 0.116 & 0.114 & 0.000 & 0.994 & 0.006 & 0.000 \\
 & Measured & 0.758 & 0.124 & 0.118 & 0.000 & 0.990 & 0.010 & 0.000 \\[3pt]
\multirow{2}{*}{Sample 2} & Target & 0.531 & 0.236 & 0.233 & 0.000 & 0.973 & 0.027 & 0.000 \\
 & Measured & 0.525 & 0.241 & 0.234 & 0.000 & 0.968 & 0.032 & 0.000 \\[3pt]
\multirow{2}{*}{Sample 3} & Target & 0.294 & 0.358 & 0.349 & 0.000 & 0.918 & 0.082 & 0.000 \\
 & Measured & 0.314 & 0.349 & 0.337 & 0.000 & 0.915 & 0.085 & 0.000 \\[3pt]
\multirow{2}{*}{Sample 4} & Target & 0.596 & 0.196 & 0.195 & 0.013 & 1.000 & 0.000 & 0.000 \\
 & Measured & 0.582 & 0.205 & 0.199 & 0.014 & 1.000 & 0.000 & 0.000 \\[3pt]
\multirow{2}{*}{Sample 5} & Target & 0.354 & 0.325 & 0.321 & 0.000 & 0.977 & 0.017 & 0.006 \\
 & Measured & 0.348 & 0.331 & 0.320 & 0.000 & 0.974 & 0.020 & 0.006 \\
\hline
\end{tabular}
\end{table}

\begin{table}[htbp]
\centering
\small
\caption{\textbf{Structural and magnetic properties of the samples.} The lattice parameters $a$ and $c$ from X-ray diffraction, the estimated main-phase volume fraction (Methods, Sec.~\ref{sec4.5}), the saturation magnetization $\mu_0 M$ calculated for the measured compositions with the fixed calculated cell volume of 84.71~\AA$^3$ (Methods, Sec.~\ref{sec4.4}), and the measured $\mu_0 M$ at 300~K, not corrected for secondary phases. Calculations for the measured compositions assume the same Fe and Mn partitioning between the 2c and 3g sites as the targets.}
\label{tab:experiment_properties}
\begin{tabular}{@{}lrrrrr@{}}
\hline
Sample & $a$ & $c$ & Phase & \multicolumn{2}{c}{$\mu_0 M$ (T)} \\
 & (\AA) & (\AA) & (\%) & Calculated & Measured \\
\hline
Sample 1 & 5.0004 & 3.9776 & 86.6 & 1.049 & 1.106 \\
Sample 2 & 5.0066 & 3.9858 & 95.9 & 1.159 & 1.099 \\
Sample 3 & 5.0179 & 3.9921 & 94.1 & 1.279 & 1.187 \\
Sample 4 & 5.0016 & 3.9818 & 93.2 & 1.109 & 1.099 \\
Sample 5 & 5.0117 & 3.9878 & 93.6 & 1.228 & 1.206 \\
\hline
\end{tabular}
\end{table}

\clearpage
\bibliography{sn-bibliography}

\begin{thebibliography}{10}
\expandafter\ifx\csname url\endcsname\relax
  \def\url#1{\burl{#1}}\fi
\expandafter\ifx\csname urlprefix\endcsname\relax\def\urlprefix{URL }\fi
\providecommand{\bibinfo}[2]{#2}
\providecommand{\bibtitlelink}[2]{{\hypersetup{hidelinks}\href{#1}{#2}}}
\providecommand{\eprint}[2][]{\url{#2}}
\providecommand{\doi}[1]{\url{https://doi.org/#1}}
\bibcommenthead

\bibitem{Curtarolo2013}
\bibinfo{author}{Curtarolo, S.} \emph{et~al.}
\newblock \bibtitlelink{https://doi.org/10.1038/nmat3568}{\bibinfo{title}{The
  high-throughput highway to computational materials design}.}
\newblock \emph{\bibinfo{journal}{Nat. Mater.}} \textbf{\bibinfo{volume}{12}},
  \bibinfo{pages}{191--201} (\bibinfo{year}{2013}).

\bibitem{Lookman2019}
\bibinfo{author}{Lookman, T.}, \bibinfo{author}{Balachandran, P.~V.},
  \bibinfo{author}{Xue, D.} \& \bibinfo{author}{Yuan, R.}
\newblock
  \bibtitlelink{https://doi.org/10.1038/s41524-019-0153-8}{\bibinfo{title}{Active
  learning in materials science with emphasis on adaptive sampling using
  uncertainties for targeted design}.}
\newblock \emph{\bibinfo{journal}{npj Comput. Mater.}}
  \textbf{\bibinfo{volume}{5}}, \bibinfo{pages}{21} (\bibinfo{year}{2019}).

\bibitem{SanchezLengeling2018}
\bibinfo{author}{S{\'a}nchez-Lengeling, B.} \& \bibinfo{author}{Aspuru-Guzik,
  A.}
\newblock
  \bibtitlelink{https://doi.org/10.1126/science.aat2663}{\bibinfo{title}{Inverse
  molecular design using machine learning: Generative models for matter
  engineering}.}
\newblock \emph{\bibinfo{journal}{Science}} \textbf{\bibinfo{volume}{361}},
  \bibinfo{pages}{360--365} (\bibinfo{year}{2018}).

\bibitem{Schmidt2019}
\bibinfo{author}{Schmidt, J.}, \bibinfo{author}{Marques, M. R.~G.},
  \bibinfo{author}{Botti, S.} \& \bibinfo{author}{Marques, M. A.~L.}
\newblock
  \bibtitlelink{https://doi.org/10.1038/s41524-019-0221-0}{\bibinfo{title}{Recent
  advances and applications of machine learning in solid-state materials
  science}.}
\newblock \emph{\bibinfo{journal}{npj Comput. Mater.}}
  \textbf{\bibinfo{volume}{5}}, \bibinfo{pages}{83} (\bibinfo{year}{2019}).

\bibitem{Himanen2019}
\bibinfo{author}{Himanen, L.}, \bibinfo{author}{Geurts, A.},
  \bibinfo{author}{Foster, A.~S.} \& \bibinfo{author}{Rinke, P.}
\newblock
  \bibtitlelink{https://doi.org/10.1002/advs.201900808}{\bibinfo{title}{Data-driven
  materials science: Status, challenges, and perspectives}.}
\newblock \emph{\bibinfo{journal}{Adv. Sci.}} \textbf{\bibinfo{volume}{6}},
  \bibinfo{pages}{1900808} (\bibinfo{year}{2019}).

\bibitem{Zunger2018}
\bibinfo{author}{Zunger, A.}
\newblock
  \bibtitlelink{https://doi.org/10.1038/s41570-018-0121}{\bibinfo{title}{Inverse
  design in search of materials with target functionalities}.}
\newblock \emph{\bibinfo{journal}{Nat. Rev. Chem.}}
  \textbf{\bibinfo{volume}{2}}, \bibinfo{pages}{0121} (\bibinfo{year}{2018}).

\bibitem{vonLilienfeld2009}
\bibinfo{author}{von Lilienfeld, O.~A.}
\newblock
  \bibtitlelink{https://doi.org/10.1063/1.3249969}{\bibinfo{title}{Accurate ab
  initio energy gradients in chemical compound space}.}
\newblock \emph{\bibinfo{journal}{J. Chem. Phys.}}
  \textbf{\bibinfo{volume}{131}}, \bibinfo{pages}{164102}
  (\bibinfo{year}{2009}).

\bibitem{vonRudorff2020}
\bibinfo{author}{von Rudorff, G.~F.} \& \bibinfo{author}{von Lilienfeld, O.~A.}
\newblock
  \bibtitlelink{https://doi.org/10.1103/PhysRevResearch.2.023220}{\bibinfo{title}{Alchemical
  perturbation density functional theory}.}
\newblock \emph{\bibinfo{journal}{Phys. Rev. Research}}
  \textbf{\bibinfo{volume}{2}}, \bibinfo{pages}{023220} (\bibinfo{year}{2020}).

\bibitem{Tahmassebpur2025}
\bibinfo{author}{Tahmassebpur, J.} \emph{et~al.}
\newblock
  \bibtitlelink{https://arxiv.org/abs/2509.07180}{\bibinfo{title}{Effective
  atom theory: Gradient-driven ab initio materials design}}.
\newblock \bibinfo{eprint}{Preprint at \url{https://arxiv.org/abs/2509.07180}}
  (\bibinfo{year}{2025}).

\bibitem{Baydin2018}
\bibinfo{author}{Baydin, A.~G.}, \bibinfo{author}{Pearlmutter, B.~A.},
  \bibinfo{author}{Radul, A.~A.} \& \bibinfo{author}{Siskind, J.~M.}
\newblock
  \bibtitlelink{https://jmlr.org/papers/v18/17-468.html}{\bibinfo{title}{Automatic
  differentiation in machine learning: a survey}.}
\newblock \emph{\bibinfo{journal}{J. Mach. Learn. Res.}}
  \textbf{\bibinfo{volume}{18}}, \bibinfo{pages}{1--43} (\bibinfo{year}{2018}).

\bibitem{Griewank2008}
\bibinfo{author}{Griewank, A.} \& \bibinfo{author}{Walther, A.}
\newblock
  \bibtitlelink{https://doi.org/10.1137/1.9780898717761}{\emph{\bibinfo{title}{Evaluating
  Derivatives: Principles and Techniques of Algorithmic Differentiation}}}
  \bibinfo{edition}{2nd} edn (\bibinfo{publisher}{SIAM},
  \bibinfo{address}{Philadelphia}, \bibinfo{year}{2008}).

\bibitem{Inui2023}
\bibinfo{author}{Inui, K.} \& \bibinfo{author}{Motome, Y.}
\newblock
  \bibtitlelink{https://doi.org/10.1038/s42005-023-01132-0}{\bibinfo{title}{Inverse
  {H}amiltonian design by automatic differentiation}.}
\newblock \emph{\bibinfo{journal}{Commun. Phys.}} \textbf{\bibinfo{volume}{6}},
  \bibinfo{pages}{37} (\bibinfo{year}{2023}).

\bibitem{Inui2024}
\bibinfo{author}{Inui, K.} \& \bibinfo{author}{Motome, Y.}
\newblock
  \bibtitlelink{https://doi.org/10.1103/PhysRevResearch.6.033080}{\bibinfo{title}{Inverse
  {H}amiltonian design of highly entangled quantum systems}.}
\newblock \emph{\bibinfo{journal}{Phys. Rev. Res.}}
  \textbf{\bibinfo{volume}{6}}, \bibinfo{pages}{033080} (\bibinfo{year}{2024}).

\bibitem{Hirasaki2024}
\bibinfo{author}{Hirasaki, Y.}, \bibinfo{author}{Inui, K.} \&
  \bibinfo{author}{Saitoh, E.}
\newblock
  \bibtitlelink{https://doi.org/10.1103/PhysRevB.110.214201}{\bibinfo{title}{Inverse
  magnetoconductance design by automatic differentiation}.}
\newblock \emph{\bibinfo{journal}{Phys. Rev. B}}
  \textbf{\bibinfo{volume}{110}}, \bibinfo{pages}{214201}
  (\bibinfo{year}{2024}).

\bibitem{Liu2023}
\bibinfo{author}{Liu, H.} \emph{et~al.}
\newblock
  \bibtitlelink{https://doi.org/10.1038/s41524-023-01080-x}{\bibinfo{title}{End-to-end
  differentiability and tensor processing unit computing to accelerate
  materials' inverse design}.}
\newblock \emph{\bibinfo{journal}{npj Comput. Mater.}}
  \textbf{\bibinfo{volume}{9}}, \bibinfo{pages}{121} (\bibinfo{year}{2023}).

\bibitem{Nam2025}
\bibinfo{author}{Nam, J.}, \bibinfo{author}{Peng, J.} \&
  \bibinfo{author}{G{\'o}mez-Bombarelli, R.}
\newblock
  \bibtitlelink{https://doi.org/10.1038/s41467-025-59543-2}{\bibinfo{title}{Interpolation
  and differentiation of alchemical degrees of freedom in machine learning
  interatomic potentials}.}
\newblock \emph{\bibinfo{journal}{Nat. Commun.}} \textbf{\bibinfo{volume}{16}},
  \bibinfo{pages}{4350} (\bibinfo{year}{2025}).

\bibitem{TamayoMendoza2018}
\bibinfo{author}{Tamayo-Mendoza, T.}, \bibinfo{author}{Kreisbeck, C.},
  \bibinfo{author}{Lindh, R.} \& \bibinfo{author}{Aspuru-Guzik, A.}
\newblock
  \bibtitlelink{https://doi.org/10.1021/acscentsci.7b00586}{\bibinfo{title}{Automatic
  differentiation in quantum chemistry with applications to fully variational
  {Hartree--Fock}}.}
\newblock \emph{\bibinfo{journal}{ACS Cent. Sci.}}
  \textbf{\bibinfo{volume}{4}}, \bibinfo{pages}{559--566}
  (\bibinfo{year}{2018}).

\bibitem{Kasim2021}
\bibinfo{author}{Kasim, M.~F.} \& \bibinfo{author}{Vinko, S.~M.}
\newblock
  \bibtitlelink{https://doi.org/10.1103/PhysRevLett.127.126403}{\bibinfo{title}{Learning
  the exchange-correlation functional from nature with fully differentiable
  density functional theory}.}
\newblock \emph{\bibinfo{journal}{Phys. Rev. Lett.}}
  \textbf{\bibinfo{volume}{127}}, \bibinfo{pages}{126403}
  (\bibinfo{year}{2021}).

\bibitem{Kasim2022}
\bibinfo{author}{Kasim, M.~F.}, \bibinfo{author}{Lehtola, S.} \&
  \bibinfo{author}{Vinko, S.~M.}
\newblock
  \bibtitlelink{https://doi.org/10.1063/5.0076202}{\bibinfo{title}{{DQC}: A
  {Python} program package for differentiable quantum chemistry}.}
\newblock \emph{\bibinfo{journal}{J. Chem. Phys.}}
  \textbf{\bibinfo{volume}{156}}, \bibinfo{pages}{084801}
  (\bibinfo{year}{2022}).

\bibitem{Zhang2022}
\bibinfo{author}{Zhang, X.} \& \bibinfo{author}{Chan, G. K.-L.}
\newblock
  \bibtitlelink{https://doi.org/10.1063/5.0118200}{\bibinfo{title}{Differentiable
  quantum chemistry with {PySCF} for molecules and materials at the mean-field
  level and beyond}.}
\newblock \emph{\bibinfo{journal}{J. Chem. Phys.}}
  \textbf{\bibinfo{volume}{157}}, \bibinfo{pages}{204801}
  (\bibinfo{year}{2022}).

\bibitem{Schmitz2026}
\bibinfo{author}{Schmitz, N.~F.}, \bibinfo{author}{Ploumhans, B.} \&
  \bibinfo{author}{Herbst, M.~F.}
\newblock
  \bibtitlelink{https://doi.org/10.1038/s41524-025-01880-3}{\bibinfo{title}{Algorithmic
  differentiation for plane-wave {DFT}: materials design, error control and
  learning model parameters}.}
\newblock \emph{\bibinfo{journal}{npj Comput. Mater.}}
  \textbf{\bibinfo{volume}{12}}, \bibinfo{pages}{6} (\bibinfo{year}{2026}).

\bibitem{Zunger1990}
\bibinfo{author}{Zunger, A.}, \bibinfo{author}{Wei, S.-H.},
  \bibinfo{author}{Ferreira, L.~G.} \& \bibinfo{author}{Bernard, J.~E.}
\newblock
  \bibtitlelink{https://doi.org/10.1103/PhysRevLett.65.353}{\bibinfo{title}{Special
  quasirandom structures}.}
\newblock \emph{\bibinfo{journal}{Phys. Rev. Lett.}}
  \textbf{\bibinfo{volume}{65}}, \bibinfo{pages}{353--356}
  (\bibinfo{year}{1990}).

\bibitem{Nordheim1931}
\bibinfo{author}{Nordheim, L.}
\newblock
  \bibtitlelink{https://doi.org/10.1002/andp.19314010507}{\bibinfo{title}{Zur
  {Elektronentheorie} der {Metalle}. {I}}.}
\newblock \emph{\bibinfo{journal}{Ann. Phys.}} \textbf{\bibinfo{volume}{401}},
  \bibinfo{pages}{607--640} (\bibinfo{year}{1931}).

\bibitem{Bellaiche2000}
\bibinfo{author}{Bellaiche, L.} \& \bibinfo{author}{Vanderbilt, D.}
\newblock
  \bibtitlelink{https://doi.org/10.1103/PhysRevB.61.7877}{\bibinfo{title}{Virtual
  crystal approximation revisited: {Application} to dielectric and
  piezoelectric properties of perovskites}.}
\newblock \emph{\bibinfo{journal}{Phys. Rev. B}} \textbf{\bibinfo{volume}{61}},
  \bibinfo{pages}{7877--7882} (\bibinfo{year}{2000}).

\bibitem{Korringa1947}
\bibinfo{author}{Korringa, J.}
\newblock
  \bibtitlelink{https://doi.org/10.1016/0031-8914(47)90013-X}{\bibinfo{title}{On
  the calculation of the energy of a {Bloch} wave in a metal}.}
\newblock \emph{\bibinfo{journal}{Physica}} \textbf{\bibinfo{volume}{13}},
  \bibinfo{pages}{392--400} (\bibinfo{year}{1947}).

\bibitem{Kohn1954}
\bibinfo{author}{Kohn, W.} \& \bibinfo{author}{Rostoker, N.}
\newblock
  \bibtitlelink{https://doi.org/10.1103/PhysRev.94.1111}{\bibinfo{title}{Solution
  of the {Schr\"{o}dinger} equation in periodic lattices with an application to
  metallic {Lithium}}.}
\newblock \emph{\bibinfo{journal}{Phys. Rev.}} \textbf{\bibinfo{volume}{94}},
  \bibinfo{pages}{1111--1120} (\bibinfo{year}{1954}).

\bibitem{Soven1967}
\bibinfo{author}{Soven, P.}
\newblock
  \bibtitlelink{https://doi.org/10.1103/PhysRev.156.809}{\bibinfo{title}{Coherent-potential
  model of substitutional disordered alloys}.}
\newblock \emph{\bibinfo{journal}{Phys. Rev.}} \textbf{\bibinfo{volume}{156}},
  \bibinfo{pages}{809--813} (\bibinfo{year}{1967}).

\bibitem{Velicky1968}
\bibinfo{author}{Veli{\v{c}}k{\'y}, B.}, \bibinfo{author}{Kirkpatrick, S.} \&
  \bibinfo{author}{Ehrenreich, H.}
\newblock
  \bibtitlelink{https://doi.org/10.1103/PhysRev.175.747}{\bibinfo{title}{Single-site
  approximations in the electronic theory of simple binary alloys}.}
\newblock \emph{\bibinfo{journal}{Phys. Rev.}} \textbf{\bibinfo{volume}{175}},
  \bibinfo{pages}{747--766} (\bibinfo{year}{1968}).

\bibitem{Gyorffy1972}
\bibinfo{author}{Gyorffy, B.~L.}
\newblock
  \bibtitlelink{https://doi.org/10.1103/PhysRevB.5.2382}{\bibinfo{title}{Coherent-potential
  approximation for a nonoverlapping-muffin-tin-potential model of random
  substitutional alloys}.}
\newblock \emph{\bibinfo{journal}{Phys. Rev. B}} \textbf{\bibinfo{volume}{5}},
  \bibinfo{pages}{2382--2384} (\bibinfo{year}{1972}).

\bibitem{Akai1989}
\bibinfo{author}{Akai, H.}
\newblock
  \bibtitlelink{https://doi.org/10.1088/0953-8984/1/43/006}{\bibinfo{title}{Fast
  {K}orringa-{K}ohn-{R}ostoker coherent potential approximation and its
  application to {FCC} {N}i-{F}e systems}.}
\newblock \emph{\bibinfo{journal}{J. Phys. Condens. Matter}}
  \textbf{\bibinfo{volume}{1}}, \bibinfo{pages}{8045--8063}
  (\bibinfo{year}{1989}).

\bibitem{Ebert2011KKR}
\bibinfo{author}{Ebert, H.}, \bibinfo{author}{K{\"o}dderitzsch, D.} \&
  \bibinfo{author}{Min{\'a}r, J.}
\newblock
  \bibtitlelink{https://doi.org/10.1088/0034-4885/74/9/096501}{\bibinfo{title}{Calculating
  condensed matter properties using the {KKR-Green}'s function method---recent
  developments and applications}.}
\newblock \emph{\bibinfo{journal}{Rep. Prog. Phys.}}
  \textbf{\bibinfo{volume}{74}}, \bibinfo{pages}{096501}
  (\bibinfo{year}{2011}).

\bibitem{Faulkner2018}
\bibinfo{author}{Faulkner, J.~S.}, \bibinfo{author}{Stocks, G.~M.} \&
  \bibinfo{author}{Wang, Y.}
\newblock
  \bibtitlelink{https://doi.org/10.1088/2053-2563/aae7d8}{\emph{\bibinfo{title}{Multiple
  Scattering Theory: Electronic Structure of Solids}}}
  (\bibinfo{publisher}{IOP Publishing}, \bibinfo{address}{Bristol},
  \bibinfo{year}{2018}).

\bibitem{Stocks1978}
\bibinfo{author}{Stocks, G.~M.}, \bibinfo{author}{Temmerman, W.~M.} \&
  \bibinfo{author}{Gyorffy, B.~L.}
\newblock
  \bibtitlelink{https://doi.org/10.1103/PhysRevLett.41.339}{\bibinfo{title}{Complete
  solution of the {Korringa}-{Kohn}-{Rostoker}
  {Coherent}-{Potential}-{Approximation} equations: {Cu-Ni} alloys}.}
\newblock \emph{\bibinfo{journal}{Phys. Rev. Lett.}}
  \textbf{\bibinfo{volume}{41}}, \bibinfo{pages}{339--343}
  (\bibinfo{year}{1978}).

\bibitem{Takahashi2007}
\bibinfo{author}{Takahashi, C.}, \bibinfo{author}{Ogura, M.} \&
  \bibinfo{author}{Akai, H.}
\newblock
  \bibtitlelink{https://doi.org/10.1088/0953-8984/19/36/365233}{\bibinfo{title}{First-principles
  calculation of the {Curie} temperature {Slater--Pauling} curve}.}
\newblock \emph{\bibinfo{journal}{J. Phys. Condens. Matter}}
  \textbf{\bibinfo{volume}{19}}, \bibinfo{pages}{365233}
  (\bibinfo{year}{2007}).

\bibitem{Lowitzer2010}
\bibinfo{author}{Lowitzer, S.}, \bibinfo{author}{K{\"o}dderitzsch, D.} \&
  \bibinfo{author}{Ebert, H.}
\newblock
  \bibtitlelink{https://doi.org/10.1103/PhysRevLett.105.266604}{\bibinfo{title}{Coherent
  description of the intrinsic and extrinsic anomalous {Hall} effect in
  disordered alloys on an ab initio level}.}
\newblock \emph{\bibinfo{journal}{Phys. Rev. Lett.}}
  \textbf{\bibinfo{volume}{105}}, \bibinfo{pages}{266604}
  (\bibinfo{year}{2010}).

\bibitem{Gutfleisch2011}
\bibinfo{author}{Gutfleisch, O.} \emph{et~al.}
\newblock
  \bibtitlelink{https://doi.org/10.1002/adma.201002180}{\bibinfo{title}{Magnetic
  materials and devices for the 21st century: Stronger, lighter, and more
  energy efficient}.}
\newblock \emph{\bibinfo{journal}{Adv. Mater.}} \textbf{\bibinfo{volume}{23}},
  \bibinfo{pages}{821--842} (\bibinfo{year}{2011}).

\bibitem{Miyake2021}
\bibinfo{author}{Miyake, T.}, \bibinfo{author}{Harashima, Y.},
  \bibinfo{author}{Fukazawa, T.} \& \bibinfo{author}{Akai, H.}
\newblock
  \bibtitlelink{https://doi.org/10.1080/14686996.2021.1935314}{\bibinfo{title}{Understanding
  and optimization of hard magnetic compounds from first principles}.}
\newblock \emph{\bibinfo{journal}{Sci. Technol. Adv. Mater.}}
  \textbf{\bibinfo{volume}{22}}, \bibinfo{pages}{543--556}
  (\bibinfo{year}{2021}).

\bibitem{Feng2014}
\bibinfo{author}{Feng, L.}, \bibinfo{author}{Liu, E.~K.},
  \bibinfo{author}{Zhang, W.~X.}, \bibinfo{author}{Wang, W.~H.} \&
  \bibinfo{author}{Wu, G.~H.}
\newblock
  \bibtitlelink{https://doi.org/10.1016/j.jmmm.2013.09.054}{\bibinfo{title}{First-principles
  investigation of half-metallic ferromagnetism of half-{Heusler} compounds
  {XYZ}}.}
\newblock \emph{\bibinfo{journal}{J. Magn. Magn. Mater.}}
  \textbf{\bibinfo{volume}{351}}, \bibinfo{pages}{92--97}
  (\bibinfo{year}{2014}).

\bibitem{Slater1937}
\bibinfo{author}{Slater, J.~C.}
\newblock
  \bibtitlelink{https://doi.org/10.1063/1.1710311}{\bibinfo{title}{Electronic
  structure of alloys}.}
\newblock \emph{\bibinfo{journal}{J. Appl. Phys.}}
  \textbf{\bibinfo{volume}{8}}, \bibinfo{pages}{385--390}
  (\bibinfo{year}{1937}).

\bibitem{Pauling1938}
\bibinfo{author}{Pauling, L.}
\newblock
  \bibtitlelink{https://doi.org/10.1103/PhysRev.54.899}{\bibinfo{title}{The
  nature of the interatomic forces in metals}.}
\newblock \emph{\bibinfo{journal}{Phys. Rev.}} \textbf{\bibinfo{volume}{54}},
  \bibinfo{pages}{899--904} (\bibinfo{year}{1938}).

\bibitem{Strnat1967}
\bibinfo{author}{Strnat, K.}, \bibinfo{author}{Hoffer, G.},
  \bibinfo{author}{Olson, J.}, \bibinfo{author}{Ostertag, W.} \&
  \bibinfo{author}{Becker, J.~J.}
\newblock \bibtitlelink{https://doi.org/10.1063/1.1709459}{\bibinfo{title}{A
  family of new cobalt-base permanent magnet materials}.}
\newblock \emph{\bibinfo{journal}{J. Appl. Phys.}}
  \textbf{\bibinfo{volume}{38}}, \bibinfo{pages}{1001--1002}
  (\bibinfo{year}{1967}).

\bibitem{Patrick2019}
\bibinfo{author}{Patrick, C.~E.} \& \bibinfo{author}{Staunton, J.~B.}
\newblock
  \bibtitlelink{https://doi.org/10.1103/PhysRevMaterials.3.101401}{\bibinfo{title}{Temperature-dependent
  magnetocrystalline anisotropy of rare earth/transition metal permanent
  magnets from first principles: the light {RCo$_5$} ($r$ = {Y}, {La}--{Gd})
  intermetallics}.}
\newblock \emph{\bibinfo{journal}{Phys. Rev. Mater.}}
  \textbf{\bibinfo{volume}{3}}, \bibinfo{pages}{101401} (\bibinfo{year}{2019}).

\bibitem{Daalderop1996}
\bibinfo{author}{Daalderop, G. H.~O.}, \bibinfo{author}{Kelly, P.~J.} \&
  \bibinfo{author}{Schuurmans, M. F.~H.}
\newblock
  \bibtitlelink{https://doi.org/10.1103/PhysRevB.53.14415}{\bibinfo{title}{Magnetocrystalline
  anisotropy of {YCo$_5$} and related {RECo$_5$} compounds}.}
\newblock \emph{\bibinfo{journal}{Phys. Rev. B}} \textbf{\bibinfo{volume}{53}},
  \bibinfo{pages}{14415--14433} (\bibinfo{year}{1996}).

\bibitem{Okumura2023}
\bibinfo{author}{Okumura, H.}, \bibinfo{author}{Fukushima, T.},
  \bibinfo{author}{Akai, H.} \& \bibinfo{author}{Ogura, M.}
\newblock
  \bibtitlelink{https://doi.org/10.1016/j.ssc.2023.115257}{\bibinfo{title}{First-principles
  calculation of magnetocrystalline anisotropy of {Y(Co,Fe,Ni,Cu)$_5$} based on
  full-potential {KKR} {Green's} function method}.}
\newblock \emph{\bibinfo{journal}{Solid State Commun.}}
  \textbf{\bibinfo{volume}{373--374}}, \bibinfo{pages}{115257}
  (\bibinfo{year}{2023}).

\bibitem{Buschow1977}
\bibinfo{author}{Buschow, K. H.~J.}
\newblock
  \bibtitlelink{https://doi.org/10.1088/0034-4885/40/10/002}{\bibinfo{title}{Intermetallic
  compounds of rare-earth and 3d transition metals}.}
\newblock \emph{\bibinfo{journal}{Rep. Prog. Phys.}}
  \textbf{\bibinfo{volume}{40}}, \bibinfo{pages}{1179--1256}
  (\bibinfo{year}{1977}).

\bibitem{Strnat1990}
\bibinfo{author}{Strnat, K.~J.}
\newblock \bibtitlelink{https://doi.org/10.1109/5.56908}{\bibinfo{title}{Modern
  permanent magnets for applications in electro-technology}.}
\newblock \emph{\bibinfo{journal}{Proc. IEEE}} \textbf{\bibinfo{volume}{78}},
  \bibinfo{pages}{923--946} (\bibinfo{year}{1990}).

\bibitem{Galanakis2002}
\bibinfo{author}{Galanakis, I.}, \bibinfo{author}{Dederichs, P.~H.} \&
  \bibinfo{author}{Papanikolaou, N.}
\newblock
  \bibtitlelink{https://doi.org/10.1103/PhysRevB.66.134428}{\bibinfo{title}{Origin
  and properties of the gap in the half-ferromagnetic {Heusler} alloys}.}
\newblock \emph{\bibinfo{journal}{Phys. Rev. B}} \textbf{\bibinfo{volume}{66}},
  \bibinfo{pages}{134428} (\bibinfo{year}{2002}).

\bibitem{deGroot1983}
\bibinfo{author}{de~Groot, R.~A.}, \bibinfo{author}{Mueller, F.~M.},
  \bibinfo{author}{van Engen, P.~G.} \& \bibinfo{author}{Buschow, K. H.~J.}
\newblock
  \bibtitlelink{https://doi.org/10.1103/PhysRevLett.50.2024}{\bibinfo{title}{New
  class of materials: Half-metallic ferromagnets}.}
\newblock \emph{\bibinfo{journal}{Phys. Rev. Lett.}}
  \textbf{\bibinfo{volume}{50}}, \bibinfo{pages}{2024--2027}
  (\bibinfo{year}{1983}).

\bibitem{Friede2024}
\bibinfo{author}{Friede, M.}, \bibinfo{author}{H{\"o}lzer, C.},
  \bibinfo{author}{Ehlert, S.} \& \bibinfo{author}{Grimme, S.}
\newblock
  \bibtitlelink{https://doi.org/10.1063/5.0216715}{\bibinfo{title}{dxtb---{A}n
  efficient and fully differentiable framework for extended tight-binding}.}
\newblock \emph{\bibinfo{journal}{J. Chem. Phys.}}
  \textbf{\bibinfo{volume}{161}}, \bibinfo{pages}{062501}
  (\bibinfo{year}{2024}).

\bibitem{Faulkner1980}
\bibinfo{author}{Faulkner, J.~S.} \& \bibinfo{author}{Stocks, G.~M.}
\newblock
  \bibtitlelink{https://doi.org/10.1103/PhysRevB.21.3222}{\bibinfo{title}{Calculating
  properties with the coherent-potential approximation}.}
\newblock \emph{\bibinfo{journal}{Phys. Rev. B}} \textbf{\bibinfo{volume}{21}},
  \bibinfo{pages}{3222--3244} (\bibinfo{year}{1980}).

\bibitem{Liechtenstein1987}
\bibinfo{author}{Liechtenstein, A.~I.}, \bibinfo{author}{Katsnelson, M.~I.},
  \bibinfo{author}{Antropov, V.~P.} \& \bibinfo{author}{Gubanov, V.~A.}
\newblock
  \bibtitlelink{https://doi.org/10.1016/0304-8853(87)90721-9}{\bibinfo{title}{Local
  spin density functional approach to the theory of exchange interactions in
  ferromagnetic metals and alloys}.}
\newblock \emph{\bibinfo{journal}{J. Magn. Magn. Mater.}}
  \textbf{\bibinfo{volume}{67}}, \bibinfo{pages}{65--74}
  (\bibinfo{year}{1987}).

\bibitem{Gyorffy1985}
\bibinfo{author}{Gyorffy, B.~L.}, \bibinfo{author}{Pindor, A.~J.},
  \bibinfo{author}{Staunton, J.}, \bibinfo{author}{Stocks, G.~M.} \&
  \bibinfo{author}{Winter, H.}
\newblock
  \bibtitlelink{https://doi.org/10.1088/0305-4608/15/6/018}{\bibinfo{title}{A
  first-principles theory of ferromagnetic phase transitions in metals}.}
\newblock \emph{\bibinfo{journal}{J. Phys. F: Met. Phys.}}
  \textbf{\bibinfo{volume}{15}}, \bibinfo{pages}{1337--1386}
  (\bibinfo{year}{1985}).

\bibitem{Hohenberg1964}
\bibinfo{author}{Hohenberg, P.} \& \bibinfo{author}{Kohn, W.}
\newblock
  \bibtitlelink{https://doi.org/10.1103/PhysRev.136.B864}{\bibinfo{title}{Inhomogeneous
  electron gas}.}
\newblock \emph{\bibinfo{journal}{Phys. Rev.}} \textbf{\bibinfo{volume}{136}},
  \bibinfo{pages}{B864--B871} (\bibinfo{year}{1964}).

\bibitem{Kohn1965}
\bibinfo{author}{Kohn, W.} \& \bibinfo{author}{Sham, L.~J.}
\newblock
  \bibtitlelink{https://doi.org/10.1103/PhysRev.140.A1133}{\bibinfo{title}{Self-consistent
  equations including exchange and correlation effects}.}
\newblock \emph{\bibinfo{journal}{Phys. Rev.}} \textbf{\bibinfo{volume}{140}},
  \bibinfo{pages}{A1133--A1138} (\bibinfo{year}{1965}).

\bibitem{Andersen1975}
\bibinfo{author}{Andersen, O.~K.}
\newblock
  \bibtitlelink{https://doi.org/10.1103/PhysRevB.12.3060}{\bibinfo{title}{Linear
  methods in band theory}.}
\newblock \emph{\bibinfo{journal}{Phys. Rev. B}} \textbf{\bibinfo{volume}{12}},
  \bibinfo{pages}{3060--3083} (\bibinfo{year}{1975}).

\bibitem{Moruzzi1978}
\bibinfo{author}{Moruzzi, V.~L.}, \bibinfo{author}{Janak, J.~F.} \&
  \bibinfo{author}{Williams, A.~R.}
\newblock
  \bibtitlelink{https://www.sciencedirect.com/science/book/9780080227054}{\emph{\bibinfo{title}{Calculated
  Electronic Properties of Metals}}}  (\bibinfo{publisher}{Pergamon Press},
  \bibinfo{address}{New York}, \bibinfo{year}{1978}).

\bibitem{Paszke2019}
\bibinfo{author}{Paszke, A.} \emph{et~al.}
\newblock
  \bibtitlelink{https://papers.nips.cc/paper_files/paper/2019/hash/bdbca288fee7f92f2bfa9f7012727740-Abstract.html}{\bibinfo{title}{{PyTorch}:
  An imperative style, high-performance deep learning library}}
  (\bibinfo{year}{2019}).
\newblock \bibinfo{note}{Advances in Neural Information Processing Systems 32}.

\bibitem{Innes2019}
\bibinfo{author}{Innes, M.}
\newblock \bibtitlelink{https://arxiv.org/abs/1810.07951}{\bibinfo{title}{Don't
  unroll adjoint: Differentiating {SSA}-form programs}}.
\newblock \bibinfo{eprint}{Preprint at \url{https://arxiv.org/abs/1810.07951}}
  (\bibinfo{year}{2018}).

\bibitem{Bradbury2018}
\bibinfo{author}{Bradbury, J.} \emph{et~al.}
\newblock \bibtitlelink{https://github.com/jax-ml/jax}{\bibinfo{title}{{JAX}:
  composable transformations of {P}ython+{N}um{P}y programs}}
  (\bibinfo{year}{2018}).
\newblock \urlprefix\url{https://github.com/jax-ml/jax}.

\bibitem{Christianson1998}
\bibinfo{author}{Christianson, B.}
\newblock
  \bibtitlelink{https://doi.org/10.1080/10556789808805697}{\bibinfo{title}{Reverse
  accumulation and implicit functions}.}
\newblock \emph{\bibinfo{journal}{Optim. Methods Softw.}}
  \textbf{\bibinfo{volume}{9}}, \bibinfo{pages}{307--322}
  (\bibinfo{year}{1998}).

\bibitem{Bai2019}
\bibinfo{author}{Bai, S.}, \bibinfo{author}{Kolter, J.~Z.} \&
  \bibinfo{author}{Koltun, V.}
\newblock \bibinfo{editor}{Wallach, H.} \emph{et~al.} (eds)
  \bibtitlelink{https://proceedings.neurips.cc/paper/2019/hash/01386bd6d8e091c2ab4c7c7de644d37b-Abstract.html}{\emph{\bibinfo{title}{Deep
  equilibrium models}}.}
\newblock (eds \bibinfo{editor}{Wallach, H.} \emph{et~al.})
  \emph{\bibinfo{booktitle}{Adv. Neural Inf. Process. Syst.}},
  Vol.~\bibinfo{volume}{32} (\bibinfo{publisher}{Curran Associates, Inc.},
  \bibinfo{year}{2019}).

\bibitem{Maliyov2025}
\bibinfo{author}{Maliyov, I.}, \bibinfo{author}{Grigorev, P.} \&
  \bibinfo{author}{Swinburne, T.~D.}
\newblock
  \bibtitlelink{https://doi.org/10.1038/s41524-024-01506-0}{\bibinfo{title}{Exploring
  parameter dependence of atomic minima with implicit differentiation}.}
\newblock \emph{\bibinfo{journal}{npj Comput. Mater.}}
  \textbf{\bibinfo{volume}{11}}, \bibinfo{pages}{22} (\bibinfo{year}{2025}).

\bibitem{Blondel2022}
\bibinfo{author}{Blondel, M.} \emph{et~al.}
\newblock \bibinfo{editor}{Koyejo, S.} \emph{et~al.} (eds)
  \bibtitlelink{https://proceedings.neurips.cc/paper/2022/hash/228b9279ecf9bbafe582406850c57115-Abstract-Conference.html}{\emph{\bibinfo{title}{Efficient
  and modular implicit differentiation}}.}
\newblock (eds \bibinfo{editor}{Koyejo, S.} \emph{et~al.})
  \emph{\bibinfo{booktitle}{Adv. Neural Inf. Process. Syst.}},
  Vol.~\bibinfo{volume}{35}, \bibinfo{pages}{5230--5242}
  (\bibinfo{publisher}{Curran Associates, Inc.}, \bibinfo{year}{2022}).

\bibitem{Bottou2018}
\bibinfo{author}{Bottou, L.}, \bibinfo{author}{Curtis, F.~E.} \&
  \bibinfo{author}{Nocedal, J.}
\newblock
  \bibtitlelink{https://doi.org/10.1137/16M1080173}{\bibinfo{title}{Optimization
  methods for large-scale machine learning}.}
\newblock \emph{\bibinfo{journal}{SIAM Rev.}} \textbf{\bibinfo{volume}{60}},
  \bibinfo{pages}{223--311} (\bibinfo{year}{2018}).

\bibitem{Hessel2020}
\bibinfo{author}{Hessel, M.} \emph{et~al.}
\newblock
  \bibtitlelink{https://github.com/google-deepmind/optax}{\bibinfo{title}{Optax:
  composable gradient transformation and optimisation, in {JAX}}}
  (\bibinfo{year}{2020}).
\newblock \urlprefix\url{https://github.com/google-deepmind/optax}.

\bibitem{LacosteJulien2016}
\bibinfo{author}{Lacoste-Julien, S.}
\newblock
  \bibtitlelink{https://doi.org/10.48550/arxiv.1607.00345}{\bibinfo{title}{Convergence
  rate of {Frank--Wolfe} for non-convex objectives}}.
\newblock \bibinfo{eprint}{Preprint at \url{https://arxiv.org/abs/1607.00345}}
  (\bibinfo{year}{2016}).

\end{thebibliography}

\end{document}